\documentclass[twocolumn,aps,prd,showpacs,amsmath,amssymb,nofootinbib,nobibnotes]{revtex4-2}
\usepackage{graphicx}
\usepackage[T1]{fontenc}
\usepackage{subfigure}
\usepackage{amsmath}
\usepackage{float}
\usepackage{mathrsfs}
\usepackage{gensymb}
\usepackage[dvipsnames]{xcolor}
\usepackage{soul}
\usepackage{enumerate}

\newcommand{\f}{\left(1-\frac{2M}{\sqrt{r^2+b^2}}\right)}
\def\be{\begin{equation}}
\def\ee{\end{equation}}

\setcitestyle{numbers,square}

\definecolor{darkblue}{RGB}{0,0,203}

\begin{document}
\title{Can different black holes cast the same shadow?
\small{}
}

\author{Haroldo C. D. Lima Junior}
\email{haroldo.ufpa@gmail.com} 
\affiliation{Faculdade de F\'{\i}sica,
Universidade Federal do Par\'a, 66075-110, Bel\'em, PA, Brazil }

\author{Lu\'{\i}s C. B. Crispino}
\email{crispino@ufpa.br} 
\affiliation{Faculdade de F\'{\i}sica,
Universidade Federal do Par\'a, 66075-110, Bel\'em, PA, Brazil }

\author{Pedro V. P. Cunha}%
 \email{pvcunha@ua.pt}
\affiliation{%
Departamento de Matem\'atica da Universidade de Aveiro and Centre for Research and Development  in Mathematics and Applications (CIDMA), Campus de Santiago, 3810-183 Aveiro, Portugal 
}%

\author{Carlos A. R. Herdeiro}
\email{herdeiro@ua.pt} 
\affiliation{Departamento de Matem\'atica da Universidade de Aveiro and Centre for Research and Development  in Mathematics and Applications (CIDMA), Campus de Santiago, 3810-183 Aveiro, Portugal }

\date{February 2021}

\begin{abstract}
We consider the following question: may two different black holes (BHs) cast exactly the same shadow? In spherical symmetry, we show the necessary and sufficient condition for a static BH to be shadow-degenerate with Schwarzschild is that the \textit{dominant} photonsphere of both has the same impact parameter, when corrected for the (potentially) different redshift of comparable observers in the different spacetimes. Such shadow-degenerate geometries are classified into two classes. The first shadow-equivalent class contains metrics whose constant (areal) radius hypersurfaces are isometric to those of the Schwarzschild geometry, which is  illustrated by the Simpson and Visser (SV) metric. The second shadow-degenerate class contains spacetimes with different redshift profiles and an explicit family of metrics within this class is presented. In the stationary, axi-symmetric case, we determine a sufficient condition for the metric to be shadow degenerate with Kerr for far-away observers.  Again we provide two classes of examples. The first class contains metrics whose constant (Boyer-Lindquist-like) radius hypersurfaces are isometric to those of the Kerr geometry, which is  illustrated by a rotating generalization of the SV metric, obtained by a modified Newman-Janis algorithm. The second class of examples  pertains BHs that fail to have the standard north-south $\mathbb{Z}_2$ symmetry, but  nonetheless remain  shadow degenerate with Kerr. The latter provides a sharp illustration that the shadow is not a probe of the horizon  geometry. These examples illustrate that nonisometric BH spacetimes can cast the same shadow, albeit the lensing is generically different.
\end{abstract}

\maketitle

\section{Introduction}
Strong gravity research is undergoing a golden epoch. After one century of theoretical investigations, the last five years have started to deliver long waited data on the strong field regime of astrophysical black hole (BH) candidates. The first detection of gravitational waves from a BH binary merger~\cite{Abbott:2016blz} and the first image of an astrophysical BH resolving horizon scale structure~\cite{Akiyama:2019cqa,Akiyama:2019fyp,Akiyama:2019eap} have opened a new era in testing the true nature of astrophysical BHs and the hypothesis that these are well described by the Kerr metric~\cite{Kerr:1963ud}.

In both these types of observations a critical question to correctly interpret the data is the issue of degeneracy. How degenerate are these observables for different models? This question, moreover, is two-fold. There is the \textit{practical} issue of degeneracy, due to the observational error bars. Different models may predict different gravitational waves or BH images which, however, are indistinguishable within current data accuracy. But there is also the \textit{theoretical} issue of degeneracy. Can different models predict the same phenomenology for some (but not all) observables? For the case of gravitational waves this is reminiscent of an old question in spectral analysis: can one hear the shape of a drum~\cite{Kac:1966xd}? (See also~\cite{Volkel:2019muj}). Or, in our context, can two different BHs be isospectral? For the case of BH imaging, this is the question if two different BHs capture light in the same way, producing precisely the same silhouette. In other words, can two different BH geometries have precisely the same shadow~\cite{Falcke:1999pj}? The purpose of this paper is to investigate the latter question.

The BH shadow is not an imprint of the BH's event horizon~\cite{Cunha:2018gql}. Rather, it is determined by a set of bound null orbits, exterior to the horizon, which, in the nonspherical cases include not only light rings (LRs) but also non-planar orbits; in Ref.~\cite{Cunha:2017eoe} these were dubbed \textit{fundamental photon orbits}. In the Kerr case these orbits are known as spherical orbits~\cite{Teo}, since they span different latitudes at constant radius, in Boyer-Lindquist coordinates~\cite{Boyer:1966qh}. 

It is conceivable that different spacetime geometries can have, in an appropriate sense, equivalent fundamental photon orbits \textit{without} being isometric to one another. In fact, one could imagine extreme situations in which one of the spacetimes is not even a BH.\footnote{A rough, but not precise, imitation of the BH shadow by a dynamically robust BH mimicker was recently discussed in~\cite{Herdeiro:2021lwl}.} It is known that equilibrium BHs, in general, must have LRs~\cite{Cunha:2020azh}, and, consequently, also non-planar fundamental photon orbits~\cite{Grover:2017mhm}. But the converse is not true: spacetimes with LRs need not to have a horizon - see, $e.g.$, Ref.~\cite{Cunha:2017qtt}.

In this paper we will show such shadow-degenerate geometries indeed exist and consider explicit examples. For spherical, static, geometries, a general criterion can be established for shadow-degenerate geometries. The latter are then classified into two distinct equivalence classes. Curiously, an interesting illustration of the simplest class is provided by the \textit{ad hoc} geometry introduced by Simpson and Visser (SV)~\cite{Simpson:2018tsi}, describing a family of spacetimes that include the Schwarzschild BH, regular BHs and wormholes. In the stationary axisymmetric case we consider the question  of shadow degeneracy with the Kerr family within the class of metrics that admit separability of  the Hamilton-Jacobi (HJ) equation. We provide two qualitatively distinct classes of examples. The first one is obtained by
 using a modified Newman-Janis algorithm~\cite{Newman:1965tw}  proposed in~\cite{Azreg-Ainou:2014pra}; we construct a rotating version of the SV spacetime, which, as shown here, turns out to belong to a shadow-degenerate class including the Kerr spacetime. A second class of examples discusses BHs with the unusual feature  that they do not possess the usual north-south $\mathbb{Z}_2$ symmetry present in,  say, the Kerr family, but nonetheless can have the same shadow as the Kerr spacetime. This provides a nice illustration  of the observation  in~\cite{Cunha:2018gql} that the shadows are \textit{not} a probe of the event horizon geometry.

This paper is organized as follows. In Sec.~\ref{sec2} we discuss a general criterion for shadow-degeneracy in spherical symmetry and classify the geometries wherein it holds into two equivalence classes. Section~\ref{sec3} discusses illustrative examples of shadow-degenerate metrics for both classes. For class I, the example is the SV spacetime. We also consider the lensing in shadow-degenerate spacetimes, which is generically non-degenerate. Section~\ref{sec4} discusses shadow degeneracy for stationary geometries admitting separability of the  HJ equation. Section~\ref{Sec5} discusses illustrative examples of shadow-degenerate metrics with Kerr, in particular constructing a rotating generalization of the SV spacetime and a BH spacetime without $\mathbb{Z}_2$ symmetry,  and analysing their lensing/shadows. We present some final remarks in Sec.~\ref{sec6}. 

\section{Shadow-degeneracy in spherical symmetry}
\label{sec2}
Let us consider a spherically symmetric, asymptotically flat, static BH spacetime.\footnote{The class of metrics~\eqref{sss} may describe non-BH spacetimes as well.} Its line element can be written in the generic form:
\begin{equation}
ds^2=-V(R)A(R)dt^2+\frac{dR^2}{B(R)V(R)}+R^2d\Omega_2 \ .
\label{sss}
\end{equation}
Here, $V(R)\equiv 1-2m/R$ is the standard Schwarzschild function, where $m$ is a constant that fixes the BH horizon at $R=2m$. The constant $m$ needs \textit{not} to coincide with the ADM mass, denoted $M$. The two arbitrary radial functions $A(R),B(R)$ are positive outside the horizon, at least $C^1$ and tend asymptotically to unity:
\be
\lim_{R\rightarrow \infty} A(R),B(R)=1 \ .
\ee
 In the line element~\eqref{sss}, $d\Omega_2 \equiv \left[d\theta^2+\sin^2\theta\,d\varphi^2\right]$ is the metric on the unit round 2-sphere.

Due to the spherical symmetry of the metric~\eqref{sss}, we can restrict the motion to the equatorial plane $\theta=\pi/2$. Null geodesics with energy $E$ and angular momentum $j$ have an impact parameter 
\be
\label{azim_momentum}\lambda\equiv \frac{j}{E} \ ,
\ee
and spherical symmetry allows us to restrict to $\lambda\geqslant 0$.\footnote{The case $\lambda=0$ has no radial turning points.}

Following Ref.~\cite{Cunha:2017qtt}, we consider the Hamiltonian $\mathscr{H}=\frac{1}{2}g^{\mu\nu}p_\mu p_\nu$, to study the null geodesic flow associated to the line element~\eqref{sss}. One introduces  a potential term $\mathcal{V}(R)\leqslant 0$ such that 
\be
\label{Rdot}2\mathscr{H}= \mathcal{V} + g_{RR}\dot{R}^2 =0 \ .
\ee 
The dot denotes differentiation regarding an affine parameter. Clearly, a radial turning point $(\dot{R}=0)$ is only possible when $\mathcal{V}=0$ for null geodesics ($\mathscr{H}=0$). One can further factorize the potential $\mathcal{V}$ as
\begin{equation}
\mathcal{V}=j^2 g^{tt}\left(\frac{1}{\lambda}-\mathcal{H}\right)\left(\frac{1}{\lambda}+\mathcal{H}\right)\leqslant 0\,,
\label{Eq-potenV}
\end{equation}
which introduces the \textit{effective potential} $\mathcal{H}(R)$:
\begin{equation}
\mathcal{H}(R)=\frac{\sqrt{A(R)\,V(R)}}{R}\geqslant 0\,.
\label{hp}
\end{equation}
In spherical symmetry, the BH shadow is determined by LRs. In the effective potential~\eqref{hp} description, a LR corresponds to a critical point of $\mathcal{H}(R)$, and its impact parameter is the inverse of  $\mathcal{H}(R)$ at the LR~\cite{Cunha:2017qtt}:
\be
\mathcal{H}'(R_{\rm LR})=0 \ , \qquad \lambda_{\rm LR}=\frac{1}{\mathcal{H}(R_{\rm LR})} \ ,
\label{glr}
\ee
where prime denotes radial derivative.

The BH shadow is an observer dependent concept. Let us therefore discuss the observation setup. We consider an astrophysically relevant scenario. First, the BH is directly along the observer's line of sight. Second, the observer is localized at the same areal radius, $R_{\rm obs}$, in the Schwarzschild and non-Schwarzschild spacetimes,\footnote{This is a geometrically significant coordinate, and thus can be compared in different spacetimes.} both having the same ADM mass $M$. Finally,  the observer is sufficiently far away so that no LRs exist for  $R>R_{\textrm{obs}}$ in both spacetimes, but $R_{\textrm{obs}}$ needs not to be at infinity.\footnote{ If a LR exists for $R>R_{\textrm{obs}}$ a more complete analysis can be done, possibly featuring a BH shadow component in the opposite direction to that of the BH, from the viewpoint of the observer. However, this will not be discussed here in order to focus on a more astrophysical setup.}
The connection between the impact parameter $\lambda$ of a \textit{generic} light ray and the observation angle $\beta$ with respect to the observer-BH  line of sight is~\cite{Cunha:2016bpi}:
\begin{equation}
\lambda=\frac{\sin\beta}{\mathcal{H}\big(R_{\textrm{obs}}\big)}\,.
\end{equation}
The degeneracy condition, that is, for the shadow edge to be the same, when seen by an observer at the same areal radius $R_{\rm obs}$ in comparable spacetimes ($i.e.$, with the same ADM mass, $M$), is that the \textit{observation angle $\beta$ coincides} in both cases, for both the metric~\eqref{sss} and the Schwarzschild spacetime. This implies that the impact parameter of the shadow edge in the generic spacetime must satisfy:
\begin{equation}
\lambda_{\rm LR}= \frac{\sqrt{27}M}{\sqrt{A_{\textrm{obs}}}}\ \ ,
\label{Eq-qObs}
\end{equation}
where we have used that, for Schwarzschild, the LR impact parameter is $\lambda^{\rm Schw}_{_\textrm{\scriptsize LR}}=\sqrt{27}M $ and $A_{\textrm{obs}}\equiv A(R_{\rm obs})$. 
Hence, only for the cases wherein $A=1$ does shadow degeneracy amounts to having the same impact parameter in the Schwarzschild and in the non-Schwarzschild spacetimes, for an observer which is not at spatial infinity. In general, the different gravitational redshift at the "same" radial position in the two spacetimes must be accounted for, leading to~\eqref{Eq-qObs}.

In the next two subsections we will distinguish two different classes of shadow degenerate spacetimes (class I and II), using the results just established. It is important to remark that the fact that two non-isometric spacetimes are shadow-degenerate does not imply that the gravitational lensing is also degenerate. In fact, it is generically not. 

This will be illustrated in Sec.~\ref{sec3} with some concrete examples for the two classes of shadow-degenerate spacetimes. An interesting example of shadow, but not lensing, degeneracy can be found in~\cite{Chen:2020qyp}, albeit in a different context. 

\subsection{Class I of shadow-degenerate spacetimes}
\label{sec21}
Specializing~\eqref{glr} for \eqref{sss} yields
\begin{equation}
\lambda_{\rm LR}=\frac{R_{\rm LR}}{\sqrt{A(R_{\rm LR})V(R_{\rm LR})}} \ .
\label{Eqimpactparameter}
\end{equation}
and
\begin{equation}
R_{\rm LR}=3m+\frac{R_{\rm LR}^4}{2\lambda_{\rm LR}^2}\frac{A'(R_{\rm LR})}{A^2(R_{\rm LR})} \ ,
\label{EqLRradius}
\end{equation}
where $A'\equiv dA/dR$. The notorious feature is that \textit{regardless} of $B(R)$, if $A(R)=1$, then it holds for the background~\eqref{sss} that $m=M$ and
\begin{equation}
R_{\rm LR}=3M  \ , \qquad \lambda_{_\textrm{\scriptsize LR}}=\sqrt{27}M \ ,
\label{ps}
\end{equation}
which are precisely the Schwarzschild results. The condition $A=1$ is thus {\it sufficient} to have the same shadow as Schwarzschild, since \eqref{Eq-qObs} is obeyed. Spacetimes~\eqref{sss} with $A=1$ define the equivalence \textit{Class I} of shadow degenerate spacetimes.

If the spacetime~\eqref{sss} with $A(R)=1$ but $B(R)$ non trivial describes a BH, it will have precisely the same shadow as the Schwarzschild spacetime. Such family of spacetimes is not fully isometric to Schwarzschild. But its constant (areal) radius hypersurfaces are isometric to those of Schwarzschild and thus have overlapping $R=$constant geodesics, which explains the result. This possibility will be illustrated in Sec.~\ref{sec31}.

These observations allow us to anticipate some shadow-degenerate geometries also for stationary, axially symmetric spacetimes. If the fundamental photon orbits are ``spherical'', not varying in some appropriate radial coordinate, as for Kerr in Boyer-Lindquist coordinates, any geometry with isometric constant radial hypersurfaces will, as in the static case, possess the same fundamental photon orbits, and will be shadow-degenerate with Kerr. We shall confirm this expectation in an illustrative example in Sec.~\ref{Sec5}.

\subsection{Class II of shadow-degenerate spacetimes}
\label{sec22}
If $A(R)\neq 1$ the solution(s) of ~\eqref{glr} will not coincide, in general, with~\eqref{ps}. In particular, 
there can be multiple critical points of $\mathcal{H}(R)$, $i.e.$ multiple LRs around the BH. This raises the question: if multiple LRs exist, which will be the one to determine the BH shadow edge? 

We define the {\it dominant} LR as the one that determines the BH shadow edge.\footnote{See~\cite{Konoplya:2020hyk, Luiz:2018prd} for a related discussion.} To determine the dominant LR first observe that:
\begin{enumerate}[(i)]
\item {Since $g_{tt}<0$ outside the horizon, the condition $1/\lambda \geqslant \mathcal{H}(R)$ must be satisfied along the geodesic motion, see Eq.~\eqref{Eq-potenV}. In particular, at LRs, the smaller the impact parameter $\lambda$ is, the larger the potential barrier $\mathcal{H}(R_{\rm LR})$ is.}
\item {The radial motion can only be inverted ($i.e.$ have a turning point) when $\mathcal{V}=0 \iff 1/\lambda=\mathcal{H}$.}
\item {The function $\mathcal{H}(R)$ vanishes at the horizon, $\left.\mathcal{H}\right|_{R=2m}=0$.}
\end{enumerate}
The BH's shadow edge is determined by critical light rays at the threshold between absorption and scattering by the BH, when starting from the observer's position $R_{\textrm{obs}}$. Considering points 1-3 above, these critical null geodesics occur at a local maximum of $\mathcal{H}$, $i.e.$, a LR. 
The infalling threshold is provided by the LR that possesses the {\it largest} value of $1/\lambda$, since this corresponds to the largest potential barrier in terms of $\mathcal{H}(R)$. Any other photonspheres that might exist with a smaller value of $1/\lambda$ do not provide the critical condition for the shadow edge, although they might play a role in the gravitational lensing. Hence, the {\it dominant photonsphere}  has the {\it smallest} value of the impact parameter $\lambda$, and shadow degeneracy with Schwarzschild is established by constraining the smallest LR impact parameter $\lambda$, via Eq.~\eqref{Eq-qObs}.

Combining Eq.~\eqref{Eq-qObs} and $ \lambda\,\mathcal{H}(R) \leqslant 1$, yields the \textit{necessary and sufficient} conditions on  $A(R)$ in order to have shadow-degeneracy with Schwarzschild. Explicitly, these are
\begin{enumerate}[(i)]
\item
\begin{equation}
A(R)\leqslant \frac{R^3}{\,\left(R-2m\right)}\,\left(\frac{A_{\textrm{obs}}}{27 M^2}\right) \, ,
\label{class2}
\end{equation}
\item the previous inequality must saturate at least once outside the horizon for some $R_{\rm LR}^{\rm dom}<R_{\textrm{obs}}$. At such dominant LR, located at $R_{\rm LR}^{\rm dom}$, \eqref{Eq-qObs} is guaranteed to hold. 
\end{enumerate}
Observe that  $3M<R_{\textrm{obs}}$ so that the observer is outside the LR in the Schwarzschild spacetime. Spacetimes~\eqref{sss} with $A\neq 1$ obeying the two conditions above define the equivalence \textit{Class II} of shadow-degenerate spacetimes.  One example will be given in Sec.~\ref{sec32}. We remark that class I of shadow-degenerate spacetimes is a particular case of class II.

\section{Illustrations of shadow-degeneracy (static)}
\label{sec3}

\subsection{Class I example: the SV spacetime}
\label{sec31}

The SV spacetime is a static, spherically symmetric geometry that generalizes the Schwarzschild metric with one additional parameter $b$, besides the ADM mass $M$, proposed in~\cite{Simpson:2018tsi}. It is an \textit{ad hoc} geometry. Its associated energy-momentum tensor, via the Einstein equations,  violates the null energy condition, and thus all classical energy conditions. Nonetheless, it is a simple and illustrative family of spacetimes that includes qualitatively different geometries. It is given by the following line element:
\begin{eqnarray}
\nonumber ds^2=&&-\f dt^2+\f^{-1}dr^2\\
\label{Lineel}&&+\left(r^2+b^2\right)d\Omega_2 \ ,
\end{eqnarray}
where the coordinates have the following domains: $
-\infty \leqslant r \leqslant \infty$, $-\infty \leqslant t \leqslant \infty$, $ 0<\theta<\pi$ and  $-\pi \leqslant \varphi<\pi$.
Depending on the value of the additional parameter $b$, which without loss of generality we assume to be non-negative, the spacetime geometry describes: (i) the Schwarzschild geometry ($b=0$); (ii) a regular BH ($0<b<2M$) with event horizon located at 
\be
r_h=\pm\sqrt{(2M)^2-b^2} \  ;
\ee
(iii) a one-way traversable wormhole geometry with a null throat at $r_t=0$ ($b=2M$); or (iv) a two-way traversable wormhole geometry with a timelike throat at $r_t=0$, belonging to the Morris-Thorne class~\cite{Morris:1988cz} ($b > 2M$).

The coordinate system in \eqref{Lineel} is relevant to observe that $r=0$ is not a singularity. Thus the geometry can be extended to negative $r$. However, it hides some other features of the geometry, since the radial coordinate in \eqref{Lineel} is not the areal radius for $b\neq 0$. Introduce the areal radius $R$ as
\be
R^2\equiv r^2+b^2 \ .
\label{ar}
\ee
The SV spacetime reads, in ($t,R,\theta,\varphi$) coordinates,
\begin{eqnarray}
 ds^2=-V(M)\,dt^2
+\frac{dR^2}{V(M)\,B_{\rm SV}(R)}
+R^2\,d\Omega_2 \ , \ \ \ \ \ \ \ 
\label{svr}
\end{eqnarray}
where
\begin{align}
&V(M)=1-\frac{2M}{R},\\
&B_{\rm SV}(R)\equiv \left(1-\frac{b^2}{R^2}\right) \ .
\label{bsv}
\end{align}
The geometry is now singular at
\be
 R= 2M\equiv R_h \ , \ \ \ {\rm and} \ \ \  R=b\equiv R_t \ .
\label{har}
\ee
For $0<b<2M$, $R_h>R_t$, the null hypersurface $R=R_h$ is a Killing horizon and the event horizon of the spacetime. It can be covered by another coordinate patch and then another coordinate singularity is found at $R=R_t$. This is again a coordinate singularity, as explicitly shown in the coordinate system~\eqref{Lineel}. It describes a throat or bounce. A free falling observer bounces back to a growing $R$, through a white hole horizon into another asymptotically flat region — see Fig. 4 in~\cite{Simpson:2018tsi}. Thus, in the coordinate system ($t,R,\theta,\varphi$), $b\leqslant R<\infty$. The other coordinate ranges are the same as before.

As it is clear from Eq.~\eqref{har}, the areal radius of the event horizon (when it exists) is $b$-independent. Moreover, since the SV spacetime is precisely of the type~\eqref{sss} with $A(R)=1$ and $B(R)= B_{\rm SV}(R)$,
it follows from the discussion of the preceding section that Eq.~\eqref{ps} holds for the SV spacetime. Thus, whenever the SV spacetime describes a BH ($0<b<2\,M$) it is class I shadow-degenerate with  a Schwarzschild BH, for an equivalent observer. This result can be also applied to the wormhole geometry if the LR is located outside the throat, i.e. the LR must be covered by the R coordinate range ($b\leq 3M$). For $b>3M$, the LR is located at the throat and Eq.~\eqref{ps} does not hold~\cite{Haroldo:2020prd}.

The LR in this spacetime has the same areal radius as in Schwarzschild, $R_{\rm LR}=3M$. However, the proper distance between the horizon and the LR, along a $t,\theta,\varphi=$constant curve is $b$-dependent:
\be
\Delta R=\int_{R_h}^{R_{\rm LR}}\sqrt{g_{RR}}dR= \left\{
 \begin{array}{l}
\simeq 3.049M \ , {\rm for} \ b=0 \ ,\\
\rightarrow \infty \ \ \ \ \ \  \ , {\rm for} \ b=2M \ .
\end{array}
\right.
\ee
It is a curious feature of the SV spacetime that the spatial sections of the horizon and the photonsphere have $b$-independent proper areas, respectively $4\pi(2M)^2$ and $4\pi (3M)^2$. But the proper distance between these surfaces is $b$-dependent and diverges as $b\rightarrow 2M$.

\begin{figure}[h!]
  \centering
  \subfigure{\includegraphics[scale=0.68]{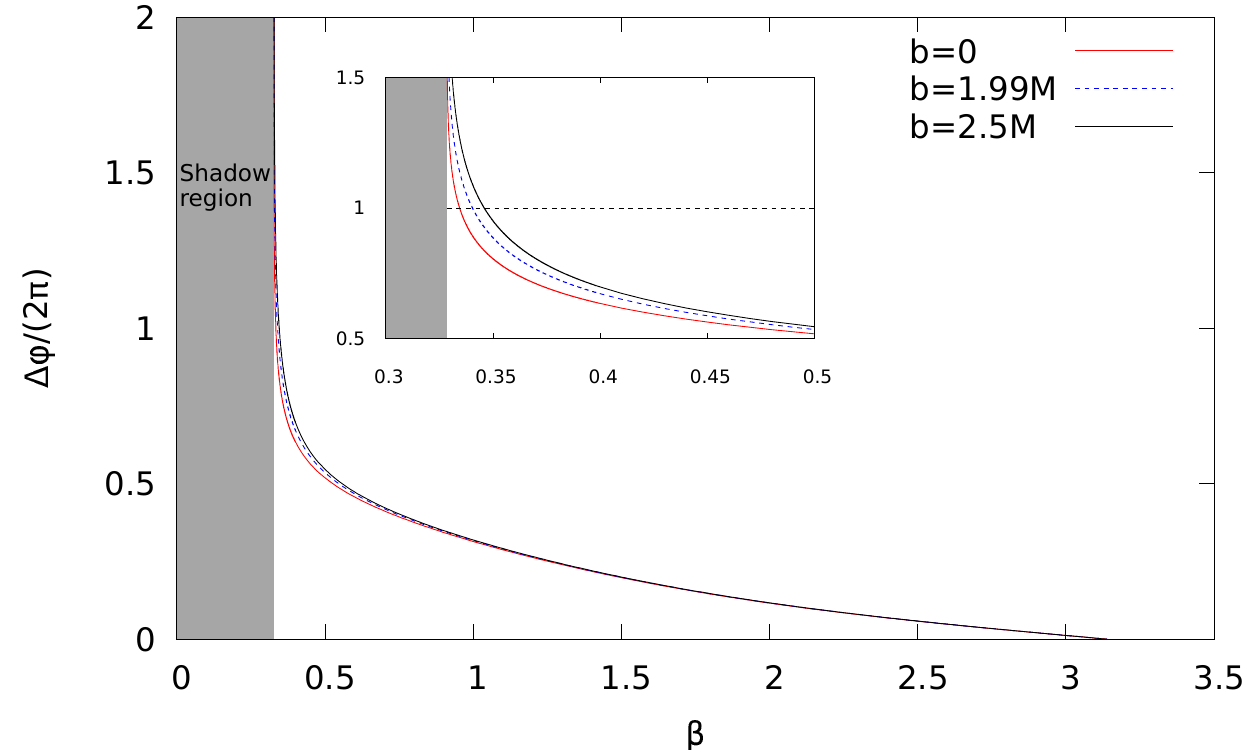}}
  \subfigure{\includegraphics[scale=0.8]{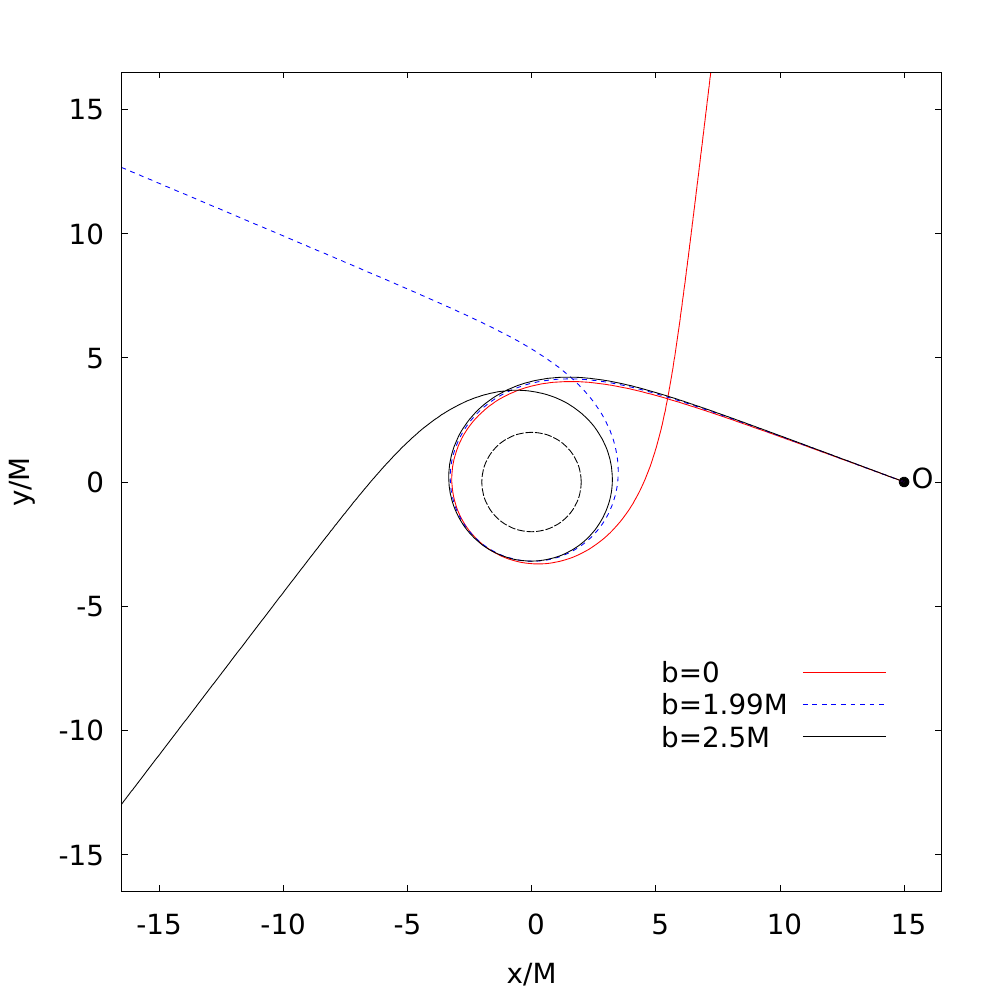}}
  \caption{Top panel: scattering angle for null geodesics as a function of the observation angle, for the Schwarzschild BH ($b=0$), SV BH ($b=1.99M$) and SV wormhole ($b=2.5M$) cases. The shaded area corresponds to the shadow region in the observer's local sky. Bottom panel: the trajectories described by null geodesics in the Cartesian plane for the same choices of $b$ and with observation angle $\beta=0.33$. The observer $O$ is located at $(x=15 M,y=0)$, as shown in the bottom panel. The dashed circle represents the event horizon of the BH cases.}
  \label{SV_null_geo}
\end{figure}

Let us now consider the gravitational lensing in the SV spacetime. We set the observer's radial coordinate equal to $R_{\rm obs}=15\,M$. In the top panel of Fig.~\ref{SV_null_geo} we plot the scattered angle $\Delta\varphi$ on the equatorial plane, in units of $2\pi$, as a function of the observation angle $\beta$. We choose three distinct values of $b$, including the Schwarzschild case ($b=0$). For large observation angles, the scattered angle is essentially the same for different values of $b$. A slight difference arises near the unstable LR, characterized by the divergent peak, as can be seen in the inset of Fig.~\ref{SV_null_geo} (top panel). In this region, for a constant $\beta$, the scattered angle increases as we increase $b$. We note that the LR, and hence the shadow region, is independent of $b$, as expected. In the bottom panel of Fig.~\ref{SV_null_geo} we show the trajectories of light rays for the observation angle $\beta=0.33$ and different values of $b$. The event horizon, for the BH cases, is represented by the dashed circle. We notice that higher values of $b$ lead to larger scattering angles.

\begin{figure}
  \centering
  \subfigure[Schwarzschild ($b=0$)]{\includegraphics[scale=0.09]{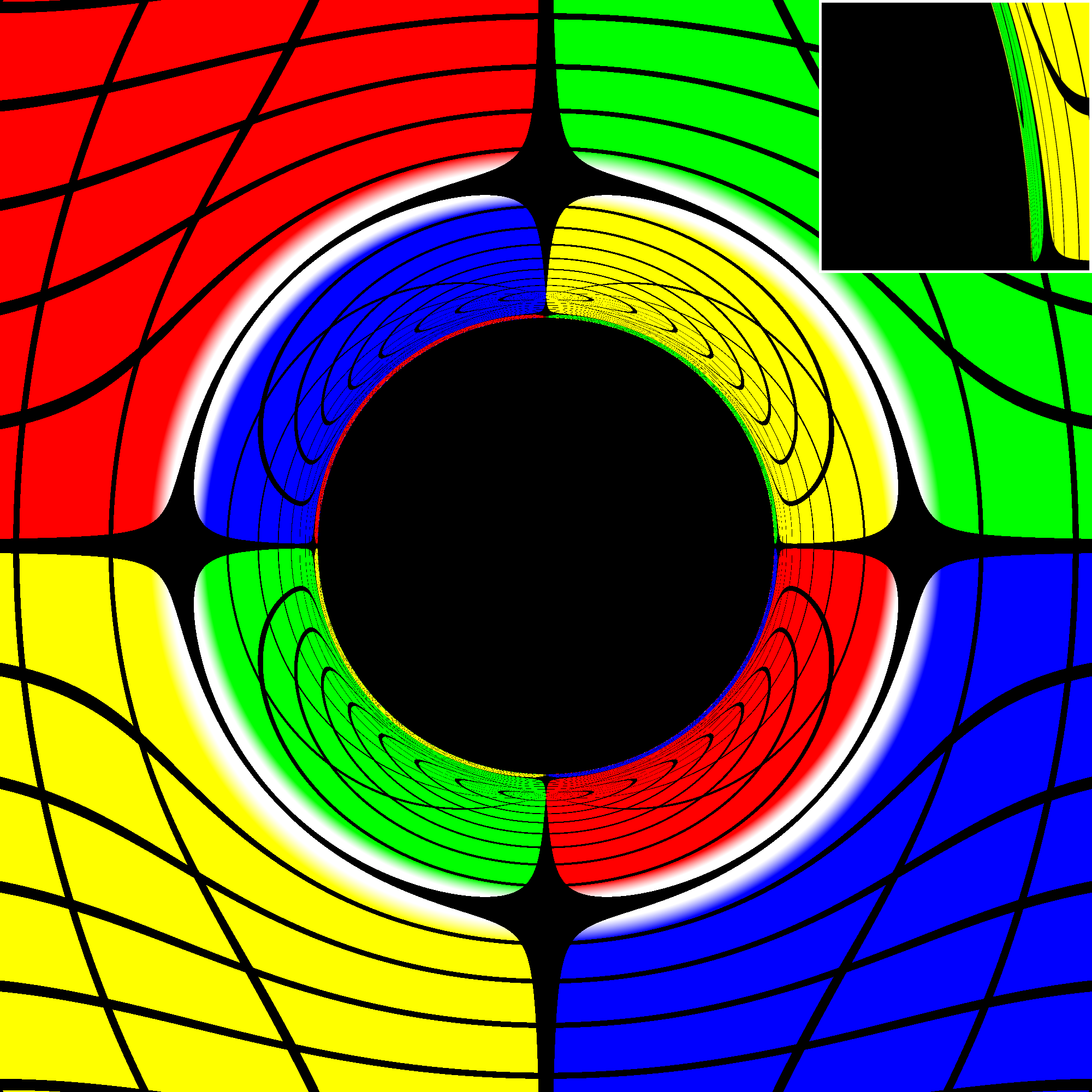}\put(-260,100){\footnotesize }}
  \subfigure[SV BH ($b=1.99M$)]{\includegraphics[scale=0.09]{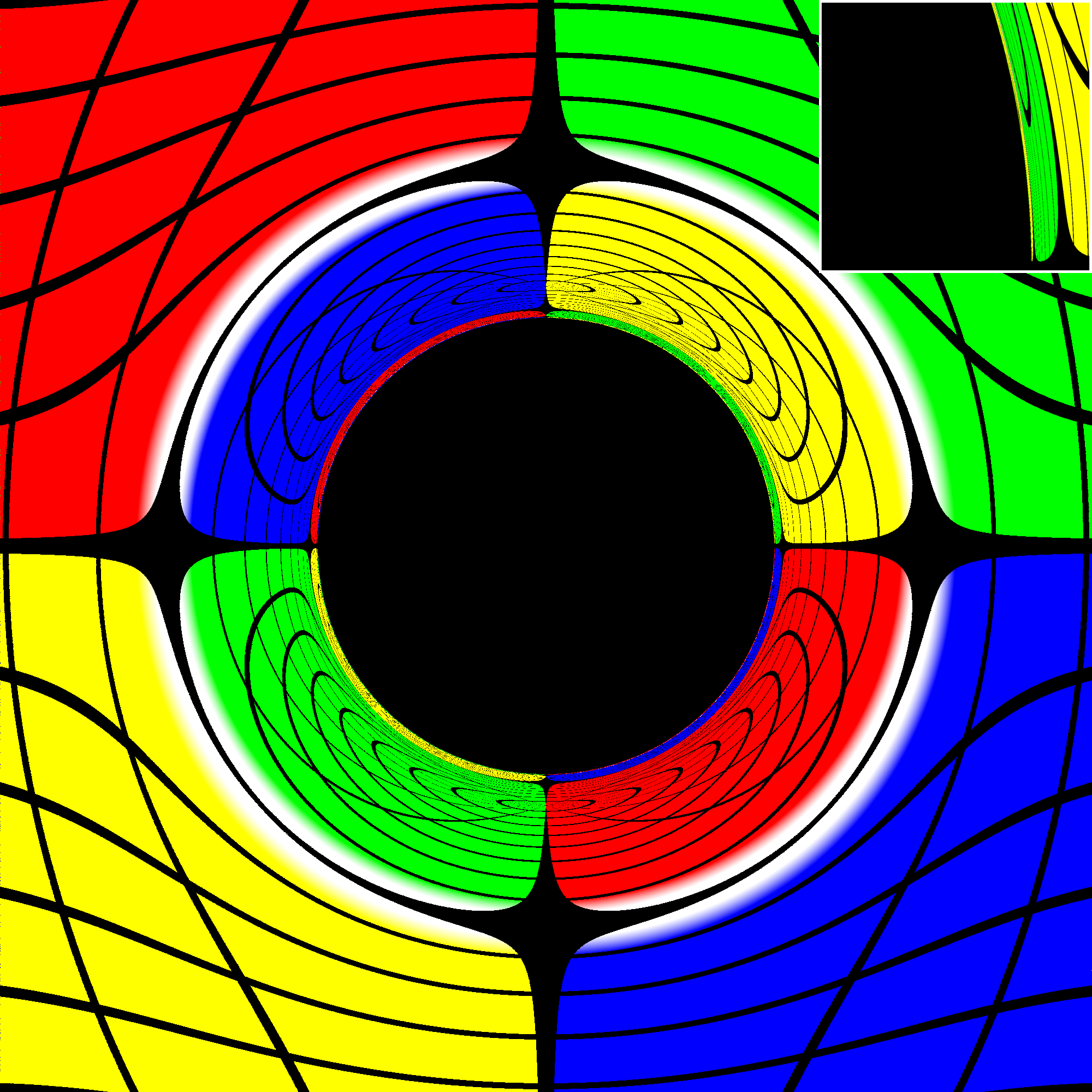}}
  \subfigure[SV wormhole ($b=2.5M$)]{\includegraphics[scale=0.09]{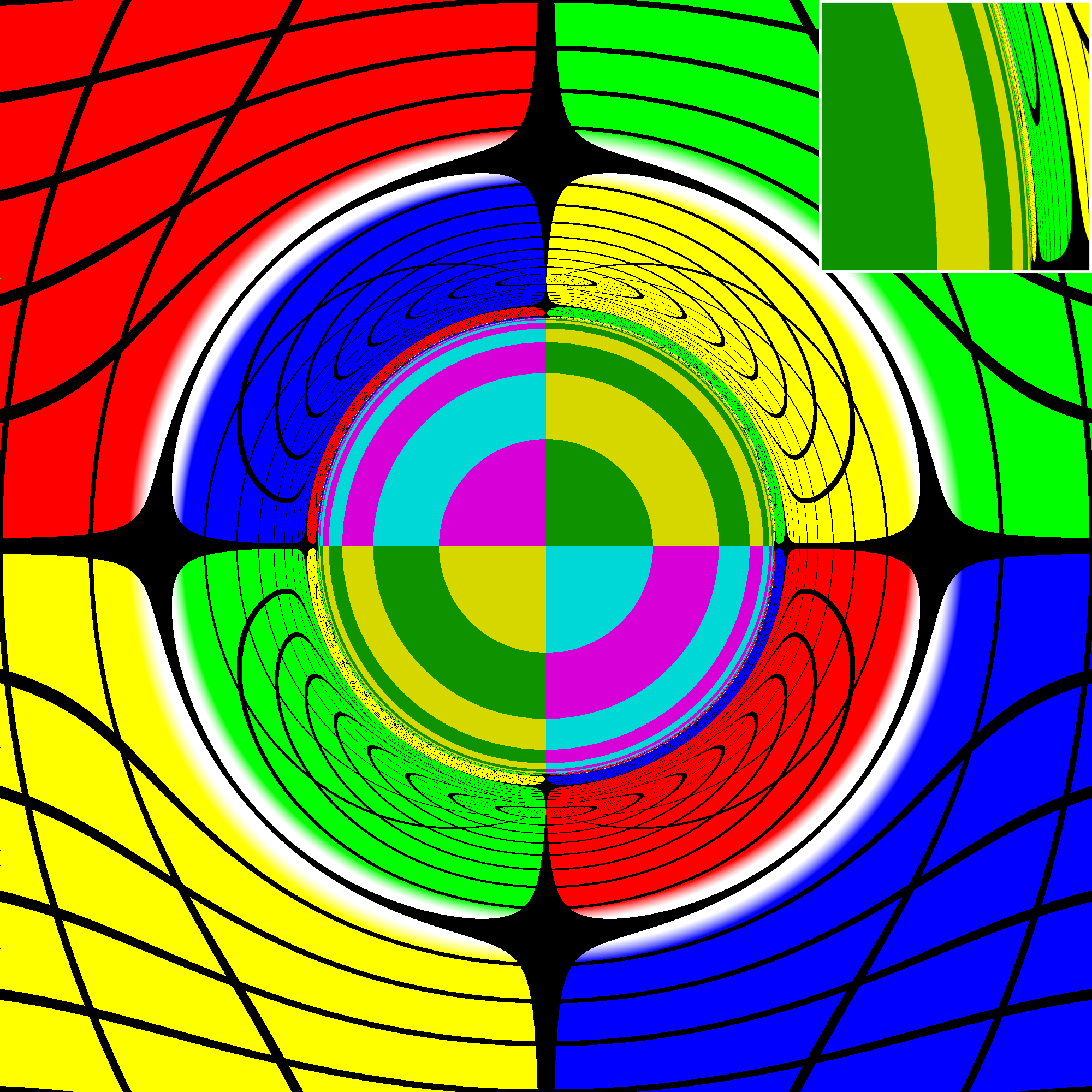}}
  \caption{The shadow and gravitational lensing of the:  Schwarzschild BH (panel a), SV BH spacetime with $b=1.99M$ (panel b), and SV wormhole spacetime with $b=2.5M$ (panel c). The observer is located on the equatorial plane and at the radial coordinate $R_{\rm obs}=15M$. The angle of view is equal to $45 \degree$. We present, in the top right corner of each panel, a zoom of the region next to the shadow edge.}
  \label{shadow_SV}
\end{figure}

We show in Fig.~\ref{shadow_SV} the shadow and gravitational lensing of the Schwarzschild BH, SV BH and SV wormhole spacetimes, obtained using backwards ray-tracing. 
In the backward ray-tracing procedure, we numerically integrate the light rays from the observer position, backwards in time, until the light rays are captured by the event horizon or scattered to infinity. {The results were obtained with two different codes: a C\texttt{++} code developed by the authors, and the \texttt{PYHOLE} code~\cite{Cunha:2016bjh} which was used as a cross-check. In this work we only show the ray-tracing results obtained with the C\texttt{++} code,  since they are essentially the same as the ones obtained with the \texttt{PYHOLE} code. }
The light rays captured by the BH  are assigned a black color in the observer's local sky. 
For the scattered light rays, we adopt a celestial sphere with four different colored quadrants (red, green, blue and yellow). A grid with constant latitude and longitude lines, and a bright spot behind the BH/wormhole are also present in the celestial sphere. This setup is similar to the ones considered in Refs.~\cite{Bohn:2014xxa, Cunha:2015yba, Cunha:2016bjh}.
{On the other hand, the light rays captured by the wormhole throat are assigned with a different color pattern (purple, cyan, dark green and dark yellow) on the celestial sphere on the other side of the throat,}
but without the grid lines and the bright spot.

The white ring surrounding the shadow region in Fig.~\ref{shadow_SV} corresponds to the lensing of the bright spot behind the BH, known as  Einstein ring~\cite{Einstein:1956zz}. It is manifest in the Schwarzschild and SV BHs, as well as in the SV wormhole case. The Einstein ring has a slightly different angular size in the three configurations shown in Fig.~\ref{shadow_SV}. This can be confirmed in Fig.~\ref{SV_null_geo}, since on the equatorial plane, the Einstein ring is formed by the light rays scattered with angle $\Delta\varphi=\pi$, which corresponds to close values of $\beta$ for the three cases. Due to the spherical symmetry, a similar analysis holds for light rays outside the equatorial plane,  which explains the formation of the Einstein ring with similar angular size in Fig.~\ref{shadow_SV}.

   Inside the Einstein ring, the whole celestial sphere appears inverted. Next to the shadow edge, there is a second Einstein ring (corresponding to a circular grid ring best seen in the inset of the figures between the yellow and green patches), in this case corresponding to the point behind the observer,  $\Delta\varphi=2\pi$ in Fig.~\ref{SV_null_geo}. A slight difference between the three configurations is also observed in this region, as can be seen in the inset of Fig.~\ref{shadow_SV}.  In between the second Einstein ring and the shadow edge there is an infinite number of copies of the celestial sphere.

\subsection{Class II example: A spacetime with $A\neq 1$}
\label{sec32}

As a concrete example of the class II of shadow-degenerate BH spacetimes with respect to the Schwarzschild BH, as seen by an observer at radial coordinate $R_{\rm obs}$, we consider (using $m=M=1$ units) the following $A(R)$ function:
\begin{equation}
\label{CII_A}A=1+\left(\frac{R_{\rm LR}-3}{b_1+b_2}\right)\!\left( a_0\left[\frac{R_{\rm LR}}{R}\right]^2  + a_1\left[\frac{R_{\rm LR}}{R}\right]^4 \right),
\end{equation}
\begin{align*}
&a_0= -2a_1 -R_{\rm LR}^3\left(1-2u^2+u^4\right),\\
&a_1= -36 + 24R_{\rm LR} - R_{\rm LR}^2 + R_{\rm LR}^3 \left(u^2 -2\right),\\
&b_1= 27 (R_{\rm LR}-2)^2,\\
&b_2= R_{\rm LR}^3 u^2 \left[ (7 - 3 R_{\rm LR})  + (2 R_{\rm LR} -5) u^2\right].
\end{align*}
We have here introduced a free parameter $R_{\rm LR}$ which sets the radial position of the dominant photonsphere (after suitably restricting the range of $R_{\rm LR}$). This spacetime is also modified by the quantity $u\equiv R_{\rm LR}/R_{\rm obs}$, which depends on the observer location. In particular, the choice $u=0$ corresponds to setting the observer at spatial infinity. This spacetime reduces to the Schwarzschild case for $R_{\rm LR}=3$ (provided than $B=1$), since it implies $A(R)=1$.
However, for $R_{\rm LR}\neq 3$ the spacetime is not Schwarzschild.\\

Not every parameter combination $\{R_{\rm LR}, R_{\rm obs}\}$ yields an acceptable spacetime for our analysis. Considering the discussion in Sec.~\ref{sec22}, the spacetime outside the horizon must satisfy both $A>0$ and Eq.~\eqref{class2}, together with $2<R_{\rm LR}<R_{\rm obs}$ and $3<R_{\rm obs}$. These conditions fix the allowed range of parameters.\\  

\begin{figure}[h!]
  \centering
  \subfigure{\includegraphics[scale=0.68]{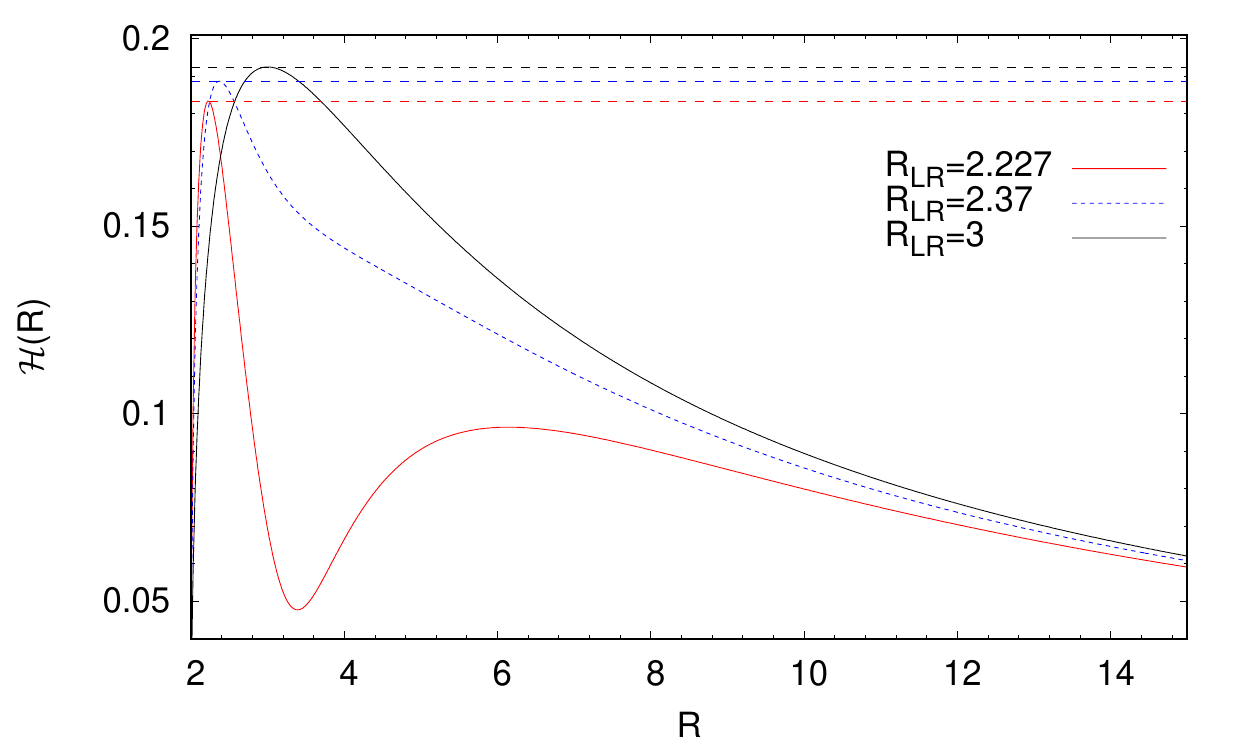}}
    \caption{The effective potential $\mathcal{H}$(R) for the class II of shadow-degenerate BHs, with $A(R)$ given by Eq.~\eqref{CII_A}, for different dominant LR radius $R_{\rm LR}$. The horizontal lines correspond to the critical impact parameters for each $R_{\rm LR}$, see Eq.~\eqref{Eq-qObs}. 
     }
  \label{effective_pot}
 \end{figure}

 For concreteness, we can set $R_{\rm obs}=15$, which leads to the following allowed range for the parameter $R_{\rm LR}$: 
\be
R_{\rm LR}\in [2.22\,,\, 3.26] \, .
\label{rangeRLR}
\ee 
In contrast, the function $B(R)$ can be left fairly unconstrained.
Curiously, for some values of $R_{\rm LR}$ in the range~\eqref{rangeRLR} there can be three photonspheres (two unstable, and one stable) outside the horizon. However, the LR at $R=R_{\rm LR}$ is always the dominant one by construction, as can be seen in Fig.~\ref{effective_pot}, where the horizontal dashed lines correspond to $1/\lambda_{\rm LR}$ - see Eq.~\eqref{Eq-qObs}. Each line intersects the maximum point of the associated potential $\mathcal{H}$, that determines the dominant LR location.

\begin{figure}
  \centering
  \subfigure{\includegraphics[scale=0.68]{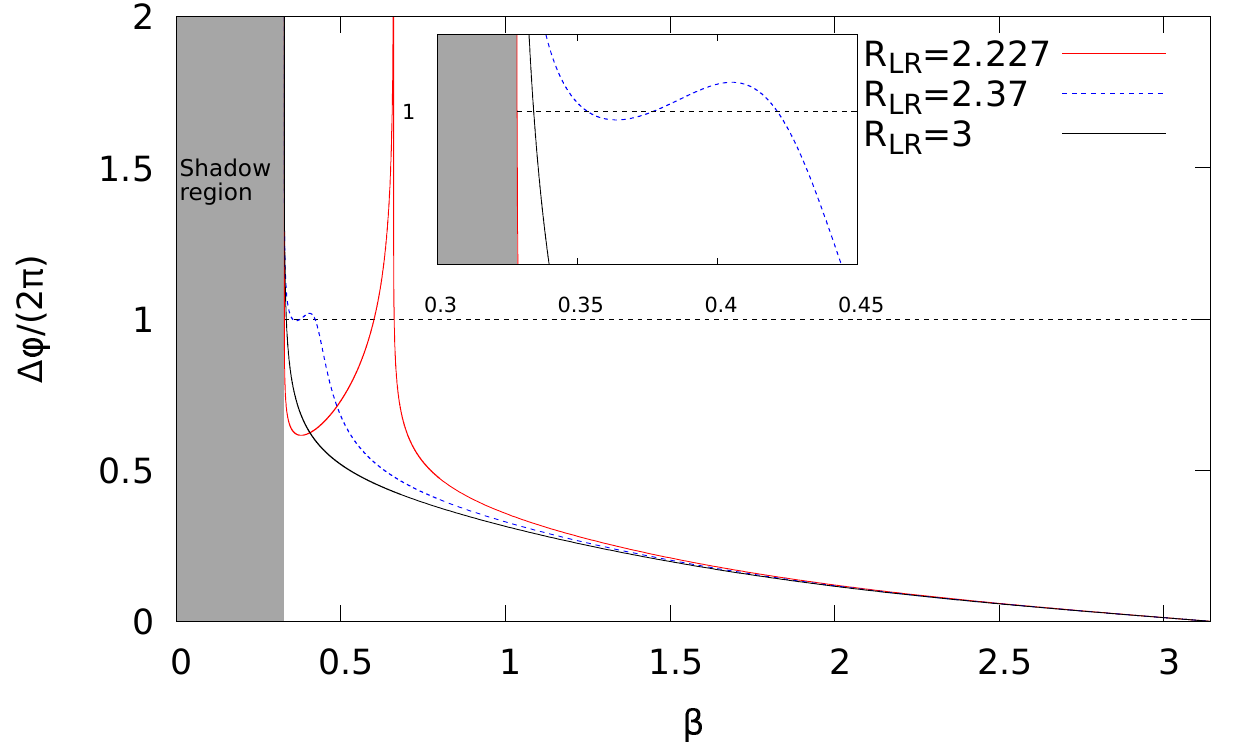}}
  \subfigure{\includegraphics[scale=0.8]{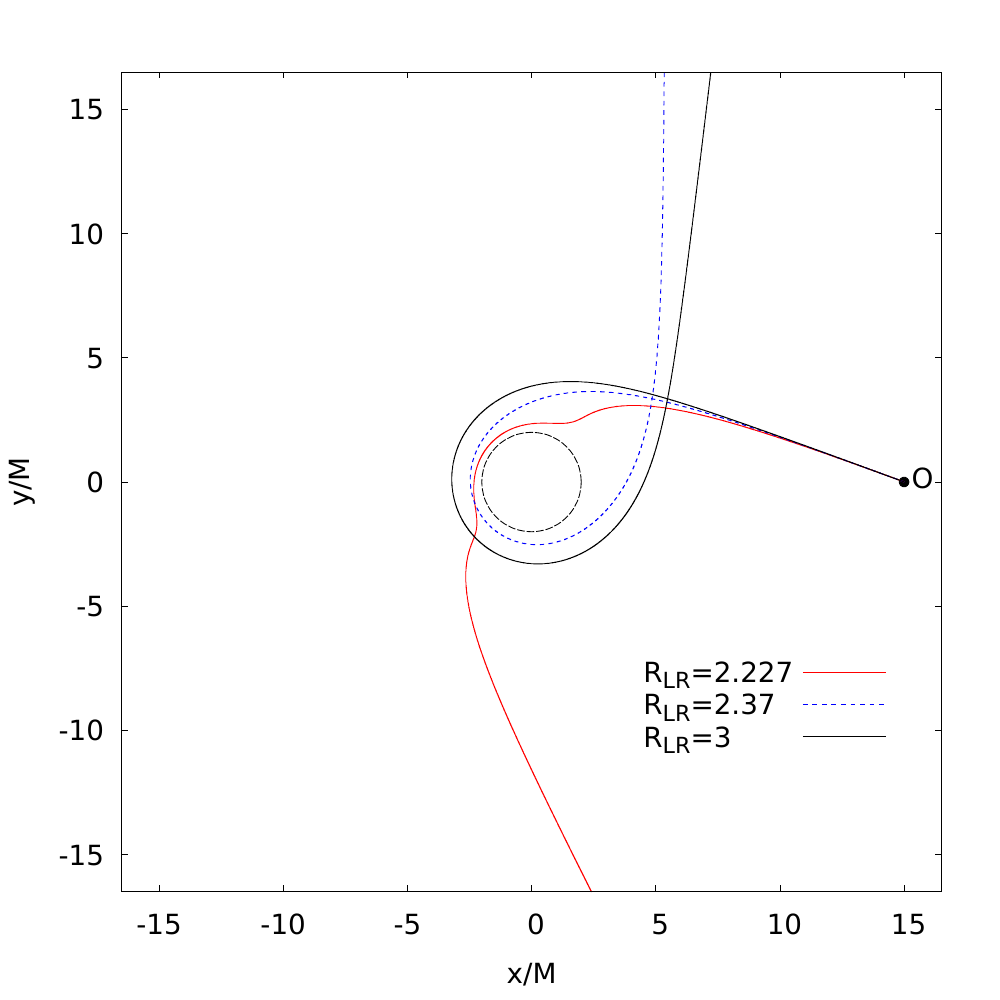}}
  \caption{Top: the scattered angle of null geodesics, as a function of the observation angle $\beta$, for $R_{\rm LR}=\lbrace 2.227, 2.37, 3 \rbrace$. The shaded area corresponds to the shadow region in the observer's local sky.  {The horizontal dotted line represents the light rays scattered at $\Delta\varphi=2\pi$ (located behind the observer). Since there is a black dot in the celestial sphere right behind the observer, the gravitational lensing displays black Einstein rings (see Fig.~\ref{shadow_CII}).} Bottom: the corresponding trajectories described by null geodesics in the equatorial plane in Cartesian-like coordinates with an observation angle $\beta=0.33$. The observer $O$ is located at $(x=15,y=0)$. The dashed circle represents the event horizon of these class II shadow degenerate examples.}
  \label{CII_lensing}
\end{figure}

Let us now consider the gravitational lensing in this spacetime, assuming for simplicity $B(R)=1$. In the top panel of Fig.~\ref{CII_lensing} we plot the scattered angle, in units of $2\pi$, as a function of the observation angle $\beta$, for the two parameter values $R_{\rm LR}=\lbrace 2.227\,,\, 2.37 \rbrace$. The Schwarzschild case $R_{\rm LR}=3$ is also included in Fig.~\ref{CII_lensing}, for reference. First taking $R_{\rm LR}=2.227$, there are two diverging peaks on the scattering angle, related to the existence of two unstable LRs (there exists also a stable LR which leaves no clear signature in the scattering plot). In contrast to the first case, $R_{\rm LR}=2.37$ contains only a single diverging peak. However, it presents a local maximum of the scattering angle next to $\beta=0.4$, as can be seen in the inset of Fig.~\ref{CII_lensing} (top panel). This local maximum leads to a duplication of lensed images. Importantly, the shadow region, corresponding to the shaded area in Fig.~\ref{CII_lensing}, is the same for the different values of $R_{\rm LR}$, as expected.\\

On the bottom panel of Fig.~\ref{CII_lensing} we show the trajectories described by null geodesics, for different values of the parameter $R_{\rm LR}$, given the same observation angle $\beta=0.33$. Curiously, when $R_{\rm LR}=2.227$, the trajectory followed by the light ray presents two inflection points when represented in the  $(x=R\cos\varphi,y=R\sin\varphi)$ plane, cf. Fig.~\ref{CII_lensing} (bottom panel). Such inflection points are determined by $d^2y/dx^2=0$, which yields

\be
\label{d2ydx2}
R^2-R\,\frac{d^2R}{d\varphi^2}+2\,\left(\frac{dR}{d\varphi}\right)^2
=0 \ \ \ \Leftrightarrow \ \ \ \ddot{R}=\frac{j^2}{R^ 3}\ ,
\ee
where the last equivalence holds for $j\neq0$.
This is also the condition for the curve $R=R(\varphi)$ to have vanishing curvature. On the other hand,  the equations of motion~\eqref{Rdot} and $\dot{\varphi}=j/R^2$, give, for $B(R)=1$,
\be
\ddot{R}=\frac{j^2}{R^ 3}+\frac{E^ 2}{2}\frac{d}{dR}\left[\frac{1}{A(R)}+\frac{2\lambda^ 2}{R^3}\right] \ .
\label{eomrdd}
\ee
Equating the last two  results yields
\be
\label{d2ydx2-cond}3+\frac{R^4}{2\lambda^2}\frac{A'(R)}{A(R)^2}=0 \ ,
\ee
as the condition at  an inflection point. Observe  the similarity with~\eqref{EqLRradius} (recall $m=1$ here); the latter has $\ddot{R}=0$, unlike~\eqref{d2ydx2}. For the Schwarzschild case, $A(R)=1$, and there are no inflection points. But for $A(R)$ given by~\eqref{CII_A} and $R_{\rm LR}=2.227$ it can be checked that the function  in square brackets in~\eqref{eomrdd} has a local maximum,  which explains the existence of inflection points.

\begin{figure}
  \centering
  \subfigure[Class II example ($R_{\rm LR}=2.227$)]{\includegraphics[scale=0.09]{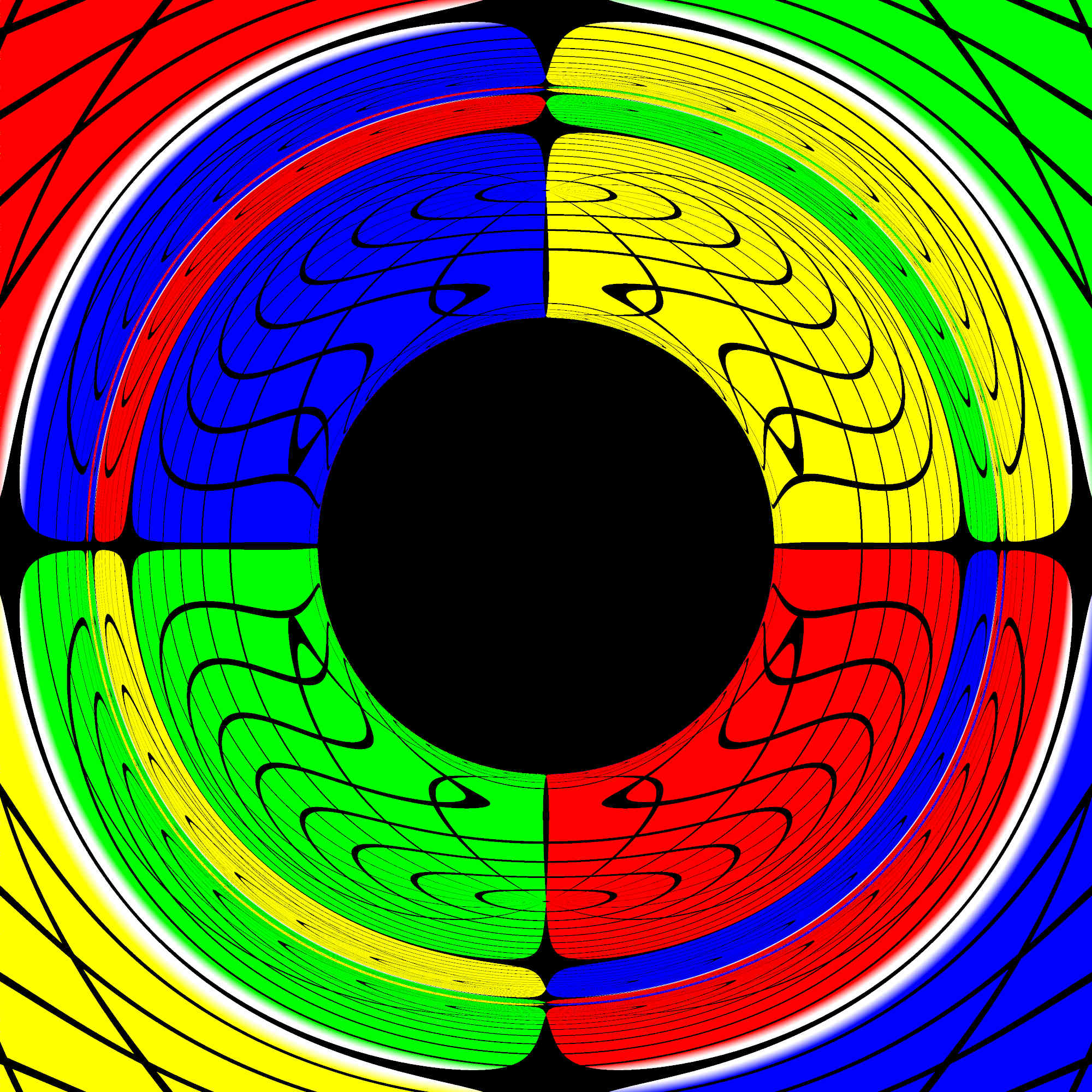}\label{shadow_CIIa}}
  \subfigure[Class II example ($R_{\rm LR}=2.37$)]{\includegraphics[scale=0.09]{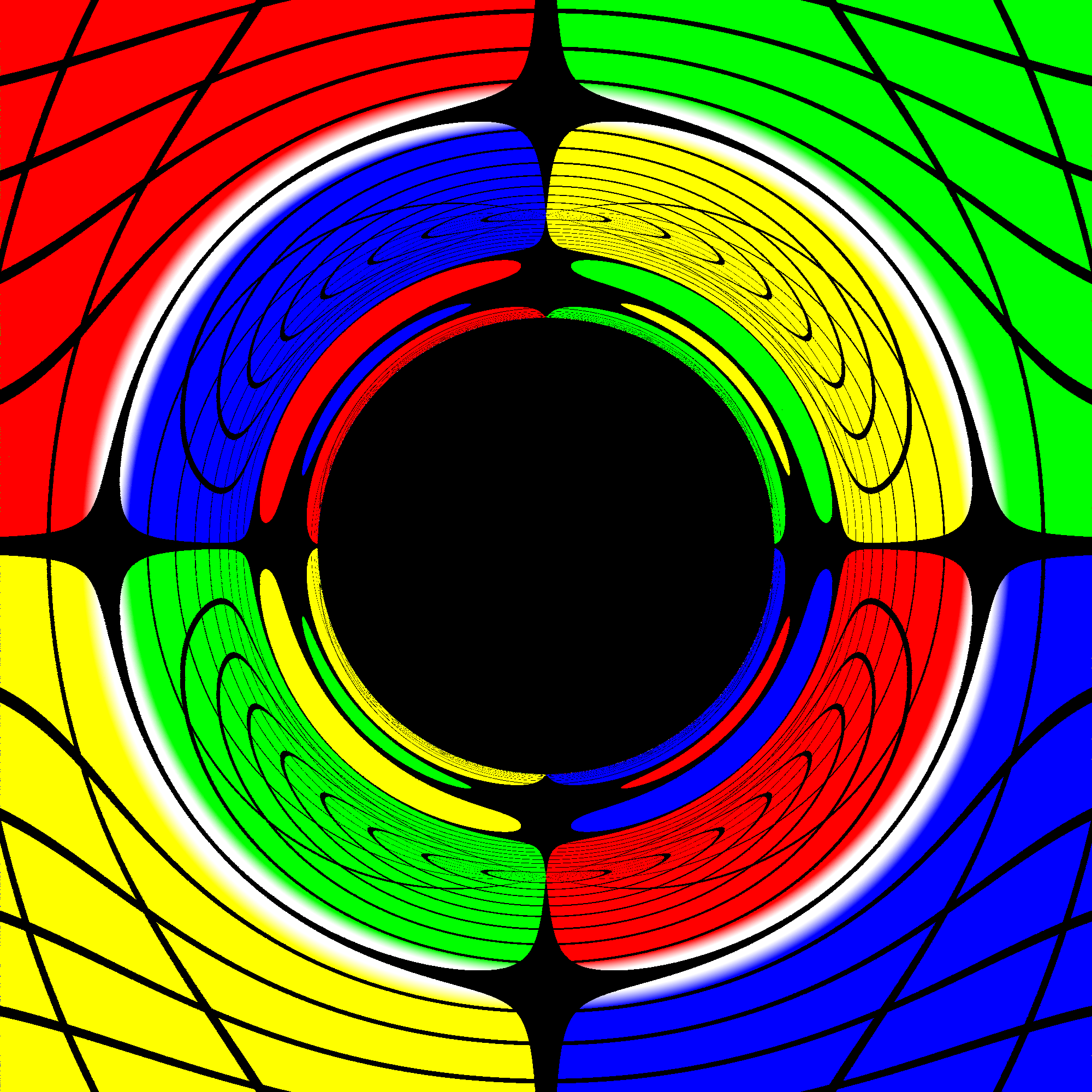}\label{shadow_CIIb}}
  \subfigure[Schwarzschild ($R_{\rm LR}=3$)]{\includegraphics[scale=0.09]{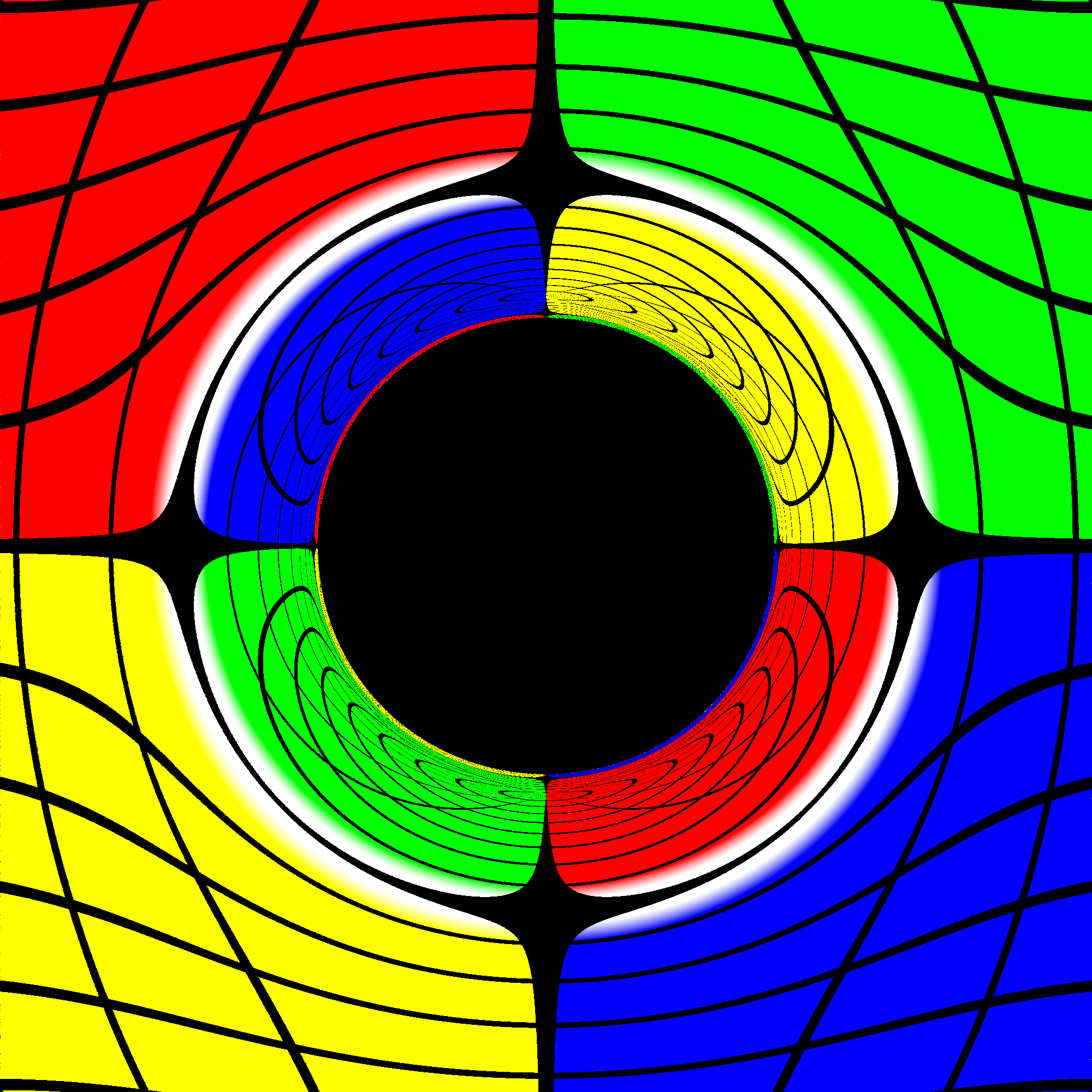}\label{shadow_CIIc}}
  \caption{The shadow and gravitational lensing for the class II example of shadow degenerate BHs, with $A(r)$ given in Eq.~\eqref{CII_A}, for $R_{\rm LR}=2.227$ (panel a), $R_{\rm LR}=2.37$ (panel b), and $R_{\rm LR}=3$ (panel c). 
  The observer position and angle of view are the same as in Fig.~\ref{shadow_SV}.}
  \label{shadow_CII}
 \end{figure}

To further illustrate the effect of different choices of the parameter $R_{\rm LR}$, we display the shadows and gravitational lensing in Fig.~\ref{shadow_CII}, obtained numerically via backwards ray-tracing. Despite identical shadow sizes, the gravitational lensing can be quite different for each value of the parameter $R_{\rm LR}$. For instance, although Einstein rings are present in all cases depicted, they have different angular diameters. This is best illustrated by looking at the white circular rings, which are mapping the point in the colored sphere directly behind the BH.

There are also some curious features of the lensing that can be anticipated from the scattering angle plot in Fig.~\ref{CII_lensing} (top panel). For example, for a parameter $R_{\rm LR}=2.227$ there are multiple copies of the celestial sphere very close to the shadow edge that are not easily identifiable in Fig.~\ref{shadow_CIIa}. This is due to light rays scattered with angles greater than $\pi$ having an observation angle $\beta$ very close to the shadow edge. The diverging peak in the scattering angle also has a clean signature in the image, in the form of a very sharp colored ring which is just a little smaller in diameter than the white circle. Additionally, taking the parameter $R_{\rm LR}= 2.37$, we can further expand our previous remark on the effect of the local maximum of the scattering angle, which introduces an image duplication of a portion of the colored sphere directly {\it behind} the observer. This feature is best seen in Fig.~\ref{shadow_CIIb} as an additional colored ring structure that does not exist in Fig.~\ref{shadow_CIIc}. \\

\section{Shadow-degeneracy in stationary BHs}
\label{sec4}
Let us now investigate possible rotating BHs with Kerr degenerate shadows. We assume that the spacetime is stationary, axi-symmetric and asymptotically flat. In contrast to the spherically symmetric case, the general motion of null geodesics in rotating BH spacetimes is not constrained to planes with constant $\theta$. This introduces an additional complication in the analysis of the geodesic motion, and in many cases it is not possible to compute the shadows analytically. For Kerr spacetime, there is an additional symmetry that allows the separability of the HJ equation, thus the shadow problem can be solved analytically. This symmetry is encoded in the so-called Carter constant, that arises due to a non-trivial Killing tensor present in the Kerr metric~\cite{Carter:1968rr}. Here, we shall investigate shadow degeneracy specializing  for rotating BH spacetimes that admit separability of the null geodesic equation.\footnote{Although this strategy implies a loss of  generality it seems unlikely (albeit a proof  is needed) that a spacetime without separability can yield precisely the same shadow as the Kerr one, which is determined  by separable fundamental photon  orbits.} A relevant previous analysis in the literature regarding shadow degeneracy for general observers can be found in~\cite{Mars:2017jkk}, albeit restricted to the Kerr-Newman-Taub-NUT BH family of solutions.

\subsection{Spacetimes admitting HJ separability}

Following  the strategy just discussed, we need the general form of the line element for a rotating BH spacetime that admits separability of the HJ equation. It is known that rotating spacetimes obtained through a Newman-Janis and modified Newman-Janis algorithms allow separability of the HJ equation~\cite{Azreg-Ainou:2014pra,Shaikh:2019prd,Junior:2020lya}. Since it is not guaranteed, however, that \textit{every} spacetime allowing separability of the HJ equation can  be  obtained by  such algorithms, we pursue a different strategy.  In Ref.~\cite{Benenti:1979}, Benenti and Francaviglia found the form of the metric tensor for a spacetime admitting separability of null geodesics in n-dimensions (see also \cite{Papadopoulos_2018}). This result is based on the following general assumptions:
\begin{enumerate}[(i)]
\item There exist (locally) $z$ independent commuting Killing vectors $\textbf{X}_\alpha$ .
\item There exist (locally) $n-z$ independent commuting Killing tensors $\textbf{K}_{a}$, satisfying
\begin{equation}
[\textbf{K}_{a},\textbf{X}_{\alpha}]=0.
\end{equation}
\item The Killing tensors $\textbf{K}_{a}$ have in common $n-z$ 
commuting 
eigenvectors $\textbf{X}_{a}$ such that
\begin{equation}
g(\textbf{X}_{a},\textbf{X}_{\alpha})=0, \qquad [\textbf{X}_{a},\textbf{X}_{\alpha}]=0.
\end{equation}
\end{enumerate}
We are interested in four-dimensional spacetimes ($n=4$) admitting two Killing vectors ($z=2$), associated to the axi-symmetry and stationarity. Hence the metric tensor that admits separability of the HJ equation is given by~\cite{Benenti:1979}:
\begin{align}
\nonumber \partial^2_s=&g^{ab}\partial_a\partial_b=\frac{1}{\tilde{A}_1(R)+\tilde{B}_1(\theta)}\left[ \left(\tilde{A}_2(R)+\tilde{B}_2(\theta)\right)\partial_t^2 \right.\\
\nonumber &+2\left(\tilde{A}_3(R)+\tilde{B}_3(\theta)\right)\partial_t\partial_\varphi+\tilde{A}_4(R)\partial_R^2+\tilde{B}_4(\theta)\partial_\theta^2\\
&\left. + \left( \tilde{A}_5(R)+\tilde{B}_5(\theta)\right) \partial_\varphi^2 \right].
\end{align}
For our purpose, it is convenient to rewrite the functions $\tilde{A}_i(R)$ and $\tilde{B}_i(\theta)$ as
\begin{align}
\nonumber &\tilde{A}_1=R^2A_1, \quad \tilde{B}_1=a^2\cos^2\theta B_1, \quad \tilde{A}_2=-\frac{(R^2+a^2)^2}{\Delta}{A_2},\\
\nonumber &\tilde{B}_2=a^2\sin^2\theta{B_2}, \quad \tilde{A}_3=-\frac{2amR}{\Delta}{A_3}, \quad \tilde{B}_3={B_3}-1,\\
\nonumber &\tilde{A}_4=\Delta{A_4}, \quad \tilde{B}_4={B_4}, \quad \tilde{A}_5=-\frac{a^2}{\Delta}{A_5}, \quad \tilde{B}_5=\frac{1}{\sin^2\theta}{B_5},
\end{align}
since we can recover the Kerr spacetime by simply taking ${A}_i=1$ and ${B}_i=1$.  The function $\Delta$ is given by
\begin{equation}
\Delta=R^2-2mR+a^2,
\end{equation}
where $m$ and $a$ are constants that fix the BH event horizon at
\begin{equation}
R_h=m+\sqrt{m^2-a^2}.
\end{equation}
Similarly to the spherically symmetric case, $m$ and $a$ need not to coincide with the ADM mass and total angular momentum per unit mass, respectively. The metric tensor in terms of ${A}_i$ and ${B}_i$ assumes the following form:
\begin{align}
\nonumber &\partial^2_s=\frac{1}{\widetilde{\Sigma}}\left[ -\left(\frac{(R^2+a^2)^2}{\Delta}A_2-a^2\sin^2\theta B_2\right)\partial_t^2 \right.\\
\nonumber &-2\left(\frac{2amR}{\Delta}A_3+1-B_3\right)\partial_t\partial_\varphi+\Delta A_4\partial_R^2+B_4\partial_\theta^2\\
\label{B&F_lineel}&\left. + \left( -\frac{a^2}{\Delta}A_5+\frac{1}{\sin^2\theta}B_5\right) \partial_\varphi^2 \vphantom{\frac{(r^2+a^2)^2}{\Delta}}\right],
\end{align}
In Eq.~\eqref{B&F_lineel}, we have
\begin{equation}
\label{sigma_tilde}\widetilde{\Sigma}=R^2 A_1+a^2\cos^2\theta B_1.
\end{equation}

We assume that $A_i(R)$ are positive outside the event horizon, and at least $C^1$. Nevertheless, the general result found by Benenti and Francaviglia may also describe non-asymptotically flat spacetimes. Hence, we need to impose constraints on the 10 functions present in the metric tensor~\eqref{B&F_lineel}, in order to describe asymptotically flat BHs. For this purpose it is sufficient that they tend asymptotically to unity, namely:
\begin{equation}
\lim_{R\rightarrow \infty}A_i(R)=1.
\end{equation}
The metric far away from the BH is given by~\cite{poisson}:
\begin{equation}
d\tilde{s}^2=-\left(1-\frac{2M}{R}\right)dt^2-\frac{2J\sin^2\theta}{R}dtd\varphi+dR^2+R^2 d\Omega_2,
\end{equation} 
where $J$ denotes the ADM angular momentum. 
{
Expanding the metric tensor \eqref{B&F_lineel} for $R\gg m$ and $R\gg a$, and comparing with the BH metric far away from the BH, we find that
\begin{align}
B_3(\theta)=B_4(\theta)=B_5(\theta)=1,
\end{align}
while $B_1(\theta)$ and $B_2(\theta)$ are left unconstrained. Hence we conclude that the metric tensor for a spacetime admitting separability and asymptotically flatness is given by
\begin{align}
\nonumber &\partial^2_s=\frac{1}{\widetilde{\Sigma}}\left[ -\left(\frac{(R^2+a^2)^2}{\Delta}A_2-a^2\sin^2\theta B_2\right)\partial_t^2 \right.\\
\nonumber &-2\frac{2amRA_3}{\Delta}\partial_t\partial_\varphi+\Delta A_4\partial_R^2+\partial_\theta^2\\
\label{gen_lineel}&\left. + \left( -\frac{a^2}{\Delta}A_5+\frac{1}{\sin^2\theta}\right) \partial_\varphi^2 \vphantom{\frac{(r^2+a^2)^2}{\Delta}}\right].
\end{align}
This line element \eqref{gen_lineel} have been obtained previously in Ref.~\cite{Chen:2020}
}

\subsection{Fundamental photon orbits}
We are now able to study the problem of shadow degeneracy in stationary and axisymmetric BHs,  using the concrete metric form \eqref{gen_lineel} for spacetimes with a  separable HJ equation. It is straightforward to compute the following geodesic equations for the coordinates $\{R,\theta\}$:
\begin{align}
&\frac{\widetilde{\Sigma}^2}{A_4}\frac{\dot{R}}{E^2}^2=\mathcal{R}(R),\\
&\widetilde{\Sigma}^2\frac{\dot{\theta}}{E^2}^2=\Theta(\theta),
\end{align}
where
\begin{align}
\nonumber &\mathcal{R}(R)=\left(R^2+a^2\right)^2A_2-4\lambda a m RA_3+a^2 \lambda^2A_5\\
\label{HJ11}&-\Delta\left(\eta+a^2+\lambda^2\right),\\
\label{HJ22}&\Theta(\theta)=\eta+a^2\left(1-\sin^2\theta B_2\right)
-\frac{\lambda^2}{\tan^2\theta}.
\end{align}
The constant parameters $\lambda$ and $\eta$ are given by
\begin{equation}
\lambda=\frac{j}{E}, \qquad\quad \eta=\frac{Q}{E^2}.
\end{equation}
As before, $\{E,j\}$ are the energy and angular momentum of the photon respectively. The quantity $Q$ is a Carter-like constant of motion, introduced via the separability of the HJ equation.

The analysis of the spherical photon orbits (SPOs) is paramount to determine the BH's shadow analytically. These orbits are a generalization of the LR orbit (photon sphere) that was discussed in the introduction. SPOs have a constant radial coordinate $R$ and are characterized by the following set of equations:
\begin{eqnarray}
\label{R0}&&\mathcal{R}(R)=0,\\
\label{DR0}&&\frac{d\,\mathcal{R}(R)}{dR}=0.
\end{eqnarray}

{In general, the set of SPOs in the spacetime \eqref{gen_lineel} will not coincide with the Kerr one. However if we set
\begin{align}
\label{shadow_deg}A_2=A_3=A_5=B_2=1,
\end{align}}%
Eq.~\eqref{HJ11} is identical to the Kerr case in Boyer-Lindquist coordinates, the same will hold for the set of equations~\eqref{R0}-\eqref{DR0}, {\textit{regardless} of $A_1$, $B_1$ and $A_4$}. One might then have the expectation that~\eqref{shadow_deg} is a sufficient condition to be shadow-degenerate with Kerr. Although this will turn out to be indeed true, caution is needed in deriving such a conclusion, since the influence of the observer frame also needs to be taken into account.

The solutions to Eqs.~\eqref{R0}-\eqref{DR0} with \eqref{shadow_deg} are the following two independent sets:
\begin{eqnarray}
\eta^{*}=-\frac{R^4}{a^2}; \ \ \lambda^*=\frac{R^2+a^2}{a},
\end{eqnarray}
and
 \begin{eqnarray}
 &&\label{crit_eta_copy}\eta=-\frac{R^3\left(R^3-6\,M\,R^2+9\,M^2\,R-4\,a^2\,M\right)}{a^2\left(R-M\right)^2},\\
 &&\label{crit_lambda_copy}\lambda=-\frac{\left(R^3-3\,M\,R^2+a^2\,R+M\,a^2\right)}{a\,\left(R-M\right)}.
 \end{eqnarray}
 The first set $\{\lambda^*,\eta^*\}$ is unphysical because~\eqref{HJ22} is not satisfied for real values. In contrast, the second set of Eqs.~\eqref{crit_eta_copy}-\eqref{crit_lambda_copy} is physically relevant: it defines the constants of motion for the different SPOs as a function of the parametric radius $R \in [R_1,R_2]$. In the latter, $R_1$ and $R_2$ are defined as the roots of Eq.~\eqref{crit_eta_copy}, given by
\begin{equation}
\frac{R_k}{M}= 2+2\cos\left\{\frac{2}{3}\arccos\left[(2k-3)\frac{\mid a\mid}{M}\right]\right\},\\
\end{equation}
where $k\in\{1,2\}$. Importantly, the physical range $R \in [R_1,R_2]$ follows from the requirement that $\Theta\geqslant 0$ (from Eq.~\eqref{HJ22}). The set of equations~\eqref{crit_eta_copy}-\eqref{crit_lambda_copy} have been extensively analyzed in the literature for the Kerr metric \cite{Teo,Wilkins:1972rs}. Remarkably, this set of orbits does not depend on the functions {$A_1$, $B_1$ and $A_4$}.\\

\subsection{Shadows}
The BH shadow edge in the spacetime \eqref{gen_lineel} is determined by the set of SPOs discussed above. To analyze the former, it is important to obtain the components of the light ray's 4-momentum  $p^\mu$, as seen by an observer in a local ZAMO (zero angular momentum observer) frame (see discussion in Ref.~\cite{Cunha:2016bpi}). In the following expressions, all quantities are computed at the observer's location: 
\begin{eqnarray}
\label{p^t}&&p^{(t)}=\sqrt{\frac{g_{\varphi\varphi}}{g^2_{t\varphi}-g_{tt}g_{\varphi\varphi}}}\left(E+\frac{g_{t\varphi}}{g_{\varphi\varphi}}j\right),\\
\label{p^r}&&p^{(R)}=\frac{p_R}{\sqrt{g_{RR}}}=\sqrt{g_{RR}}\,\dot{R},\\
\label{p^theta}&&p^{(\theta)}=\frac{p_\theta}{\sqrt{g_{\theta\theta}}}=\sqrt{g_{\theta\theta}}\,\dot{\theta},\\
\label{p^phi}&&p^{(\varphi)}=\frac{j}{\sqrt{g_{\varphi\varphi}}}.
\end{eqnarray}
One generically requires two observation angles $\{\alpha,\beta\}$, measured in the observer's frame, to fully specify the detection direction of a light ray in the local sky. These angles are defined by the ratio of the different components of the light ray's four momentum in the observer frame. Following~\cite{Cunha:2016bpi}, we can define the angles $\alpha,\beta$ to satisfy:
\begin{equation}
\label{anglesobserver}\sin\alpha=\frac{p^{(\theta)}}{p^{(t)}}\ , \qquad \tan\beta=\frac{p^{(\varphi)}}{p^{(R)}}\,.
\end{equation}
In addition, one can expect the angular size of an object to decrease like $\alpha \sim 1/R_{\rm circ}$ in the limit of far away observers, where the circumferential radius\footnote{The quantity $R_{\rm circ}$ is computed by displacing the observer to the equatorial plane ($\theta=\pi/2$), while keeping its coordinate $R$ fixed;   $R_{\rm circ}=\sqrt{g_{\varphi\varphi}}$ at that new location.} $R_{\rm circ}$ is a measure of the observer's distance to the horizon~\cite{Cunha:2016bpi,Cunha:2018acu}. Given this asymptotic behavior, it is useful to introduce the impact parameters:
\begin{equation}
X=-R_{\rm circ}\beta,\qquad Y=R_{\rm circ}\alpha\,.
\end{equation} 
By construction, these quantities $\{X,Y\}$ are expected to have a well defined limit when taking $R_{\rm circ}\rightarrow \infty$.\\
The relation between $\{X,Y\}$ and the constants of null geodesic motion $\{\lambda,\eta\}$ can be obtained by considering Eqs.~\eqref{HJ11}-\eqref{HJ22} for $\dot{R}$ and $\dot{\theta}$, and then combining them with Eqs.~\eqref{p^t}-\eqref{anglesobserver}.\\

We can compute the shadow edge expression in the limit of far-away observers ($R_{\rm circ}\rightarrow \infty$):
{
\begin{align}
&X=-\frac{\lambda}{\sin\theta_0},\\
\label{Y_parameter}&Y=\pm \sqrt{\eta+a^2\cos^2\theta_0 -\frac{\lambda^2}{\tan^2\theta_0}}\,.
\end{align}}
{The shadow degeneracy occurs when the quantities $\lbrace X,Y\rbrace$ coincide in the non-Kerr and Kerr spacetimes, for an observer located at the same circumferential radius $R_{\rm circ}$ and polar angle $\theta_0$. This will certainly be the case if the set of SPOs coincides in both Kerr and the non-Kerr geometry for observers that are very far-away. Thus recalling Eq.~\eqref{shadow_deg}, we note that the latter is a {\it sufficient} condition for shadow degeneracy at infinity.\\

We conclude with the following line element obtained from~\eqref{gen_lineel} together with ~\eqref{shadow_deg}:
\begin{align}
\nonumber &ds^2=-\frac{\left(\Delta-a^2\sin^2\theta\right)\widetilde{\Sigma}}{\Sigma^2}dt^2+\frac{\widetilde{\Sigma}}{A_4\Delta}dR^2+\widetilde{\Sigma}d\theta^2\\
\nonumber&-\frac{4amR\sin^2\theta\widetilde{\Sigma}}{\Sigma^2}dtd\phi\\
\label{DS_lineel}&+\frac{\left[\left(R^2+a^2\right)^2-a^2\Delta\sin^2\theta\right]\widetilde{\Sigma}}{\Sigma^2}\sin^2\theta d\phi^2,
\end{align}
where
\begin{align}
&\Sigma=R^2+a^2\cos^2\theta.
\end{align}
This geometry will be shadow degenerate with respect to the Kerr spacetime for very far-away observers, with very weak constraints on $A_1, B_1$ and $A_4$.
}

\section{Illustrations of shadow-degeneracy (stationary)}
\label{Sec5}

\subsection{Rotating SV spacetime}
\label{Sec51}
{An  example of a rotating, stationary and axisymmetric BH with a Kerr degenerate shadow,} 
can be obtained by applying  the method proposed in~\cite{Azreg-Ainou:2014pra} to the static SV geometry.\footnote{As this paper was being completed a  rotating version of the SV spacetime was independently constructed, using the NJA, in~\cite{Mazza:2021rgq,Shaikh:2021yux}.} This method consists on a variation of the Newman-Janis algorithm (NJA)~\cite{Newman:1965tw}. Starting from a static seed metric, we can generate via this modified NJA a rotating circular spacetime that can be always expressed in Boyer-Lindquist-like coordinates. This comes in contrast to the standard NJA, for which the latter is not always possible. However, the method introduces a new unknown function ($\Psi$ below) that may be fixed by additional (physical) arguments, for instance, interpreting the component of the stress-energy tensor $T_{\mu\nu}$ to be those of a fluid~\cite{Azreg-Ainou:2014pra}.

Applying the modified NJA discussed in Ref.~\cite{Azreg-Ainou:2014pra} to the seed metric~\eqref{Lineel} (class I spacetime - SV), we introduce the functions:
\begin{eqnarray}
&&F(r)= \f,\\
&&K(r)=r^2+b^2.
\end{eqnarray}
The latter contains the same functional structure of some of the metric elements of the seed metric~\eqref{Lineel}. Combining these functions $F,K$ with a complex change of coordinates (which introduces a parameter $a$), we obtain a possible rotating version of the SV metric~\eqref{Lineel}, namely:
\begin{eqnarray}
\nonumber &&ds^2= -\frac{\left(F K+a^2\cos^2\theta\right)}{\left(K+a^2\cos^2\theta\right)^2}\Psi dt^2+\frac{\Psi}{FK+a^2}dr^2\\
\label{rotating_SV-1} &&-2a\sin^2\theta\left[\frac{K-FK}{\left(K+a^2\cos^2\theta\right)^2}\right]\Psi dt d\varphi+\Psi d\theta^2\\
\nonumber &&+\Psi\sin^2\theta\left[1+a^2\sin^2\theta\frac{2K-FK+a^2\cos^2\theta}{\left(K+a^2\cos^2\theta\right)^2}\right]d\varphi^2,
\end{eqnarray}
where $\Psi$ is an undefined function that can be fixed by an additional requirement. Assuming a matter-source content of a rotating fluid, we can check that setting $\Psi=r^2+a^2\cos^2\theta+b^2$ leads to an Einstein tensor that satisfies the suitable fluid equations detailed in Ref.~\cite{Azreg-Ainou:2014pra}. We may then rewrite the line element \eqref{rotating_SV-1} in terms of the radial coordinate $R=\sqrt{r^2+b^2}$, which leads to the more compact form:
\begin{eqnarray}
\nonumber &&ds^2=-\left[1-\frac{2\,M\,R}{\Sigma}\right]\,dt^2+\frac{\Sigma}{\Delta}\,\frac{dR^2}{B_{\rm SV}(R)}\\
&&\nonumber-\frac{4\,M\,a\,R\,\sin^2\theta}{\Sigma}\,dt\,d\varphi+\Sigma\,d\theta^2+\\
\label{Rotating_BB_Rcord}&&+\Sigma\left[1+\frac{a^2\,\sin^2\theta}{\Sigma^2}\left(\Sigma+2\,M\,R\right)\right]\,\sin^2\theta\,d\varphi^2,
\end{eqnarray}
where 
\begin{eqnarray}
&&\Sigma=R^2+a^2\,\cos^2\theta,\\
\label{Delta}&&\Delta=R^2-2\,M\,R+a^2.
\end{eqnarray}
For simplicity, we shall designate the geometry~\eqref{Rotating_BB_Rcord} as the {\it rotating SV} spacetime.
Curiously, the line element \eqref{Rotating_BB_Rcord} is precisely the Kerr one, except for the extra factor $B_{\rm SV}(R)$ in the $g_{RR}$ component. For $a =0$, the SV line element \eqref{svr} is recovered; for $b= 0$ we recover the Kerr metric in Boyer-Lindquist coordinates.

{An important remark is in order. Depending on the coordinate choice of the seed metric~\eqref{Lineel}, the latter might be mapped to a different final geometry. For instance, had we applied the modified NJA to the seed SV metric in the coordinates of Eq.~\eqref{svr} rather than those of Eq.~\eqref{Lineel}, then we would have obtained a rotating spacetime different than that of~\eqref{Rotating_BB_Rcord}.}

Let us now examine some of the properties of the spinning SV geometry~\eqref{Rotating_BB_Rcord}.\\

\subsubsection{Singularities and horizons}
The line element \eqref{Rotating_BB_Rcord} presents singularities at:
 \begin{eqnarray}
\label{horizons1} &&R_{\pm}=M\pm \sqrt{M^2-a^2},\\
 &&R_t=b.
 \end{eqnarray}
These are coordinate singularities and the spacetime is regular everywhere. In particular~\eqref{horizons1} are Killing horizons that only exist if $R_\pm>R_t$, or
 \begin{eqnarray}
 \label{horizoncond}M\pm \sqrt{M^2-a^2} \geqslant  b.
 \end{eqnarray}
 Adopting the positive sign in Eq.~\eqref{horizoncond}, a BH only exists if $M+ \sqrt{M^2-a^2} >  b$ {and $a<M$}; for $M+ \sqrt{M^2-a^2} =  b$ the geometry describes a wormhole with a throat, { which can be nulllike, spacelike or timelike, depending on the value of $a$ and $b$}~\cite{Mazza:2021rgq}. {These singularities $R_{\pm}$ can be removed by writing the line element in Eddington-Finkelstein-like coordinates. On the other hand, the singularity $R_t$ can be removed by writing the line element in the coordinates given by Eq.~\eqref{rotating_SV-1}, in which $R_t$ corresponds to the radial coordinate $r=0$.}

In order to inquire if the geometry \eqref{Rotating_BB_Rcord} is regular everywhere, let us consider
some curvature invariants:  the Kretschmann scalar ($K=R_{\mu\,\nu\,\beta\,\rho}\,R^{\mu\,\nu\,\beta\,\rho}$), the Ricci scalar ($R^{\mu\,\nu}\,g_{\mu\,\nu}$), and the squared Ricci tensor ($R_{\mu\,\nu}\,R^{\mu\,\nu}$). The full expressions of the curvature invariants are too large to write down. {However, we 
can write them in a compact form as:
\begin{align}
&K=\frac{\mathcal{P}(R,\theta)}{2R^6\Sigma^6},\\
&R^{\mu\nu} g_{\mu\nu}=\frac{b^2\mathcal{Q}(R,\theta)}{R^3\Sigma^3},\\
&R^{\mu\nu}R_{\mu\nu}=\frac{b^4\mathcal{S}(R,\theta)}{2R^6\Sigma^6},
\end{align}
where $\mathcal{P}(R,\theta), \mathcal{Q}(R,\theta)$ and $\mathcal{S}(R,\theta)$ are polynomials of powers of $R$, $\sin\theta$ and $\cos\theta$. 
We observe that the curvature invariants are all finite {in the range of the radial coordinate $R$ ($b<R<\infty$), for $b\neq 0$. The Carter-Penrose diagram, as well as further properties of this rotating SV geometry can be found in Ref.~\cite{Mazza:2021rgq}}. 
%
%

\subsubsection{Shadow and lensing}

{The rotating SV geometry is a particular case of Eq.~\eqref{DS_lineel}, since
\begin{equation}
A_1=1, \quad B_1=1, \quad A_4=B_{SV}(R). 
\end{equation}
Hence, this BH geometry is shadow degenerate with Kerr spacetime, regardless of $b$.}
A plot of the shadow edge is presented in Fig.~\ref{figura01} (top panel) for the rotating SV spacetime with different values of $b$. As was previously mentioned, the shadow does not depend on the parameter $b$. {Notwithstanding, the dependence on $b$ through $B_{\rm SV}(R)$ has a subtle impact on the gravitational lensing, as can be seen in Fig.~\ref{shadow_NJ_SV}, where we show the shadow and gravitational lensing of the rotating SV spacetime, obtained using backwards ray-tracing.}

{We remark that applying the modified NJA to the spherically symmetric and static geometries presented in Sec.~\ref{sec2} \textit{does not} generically result in a Kerr degenerate shadow geometry. In particular, if one applies the modified NJA to the class II shadow degenerate example~\eqref{CII_A}, the resulting rotating BH geometry is not shadow degenerate.}

\begin{figure}[h!]
\subfigure[Rotating SV]{\includegraphics[scale=0.8]{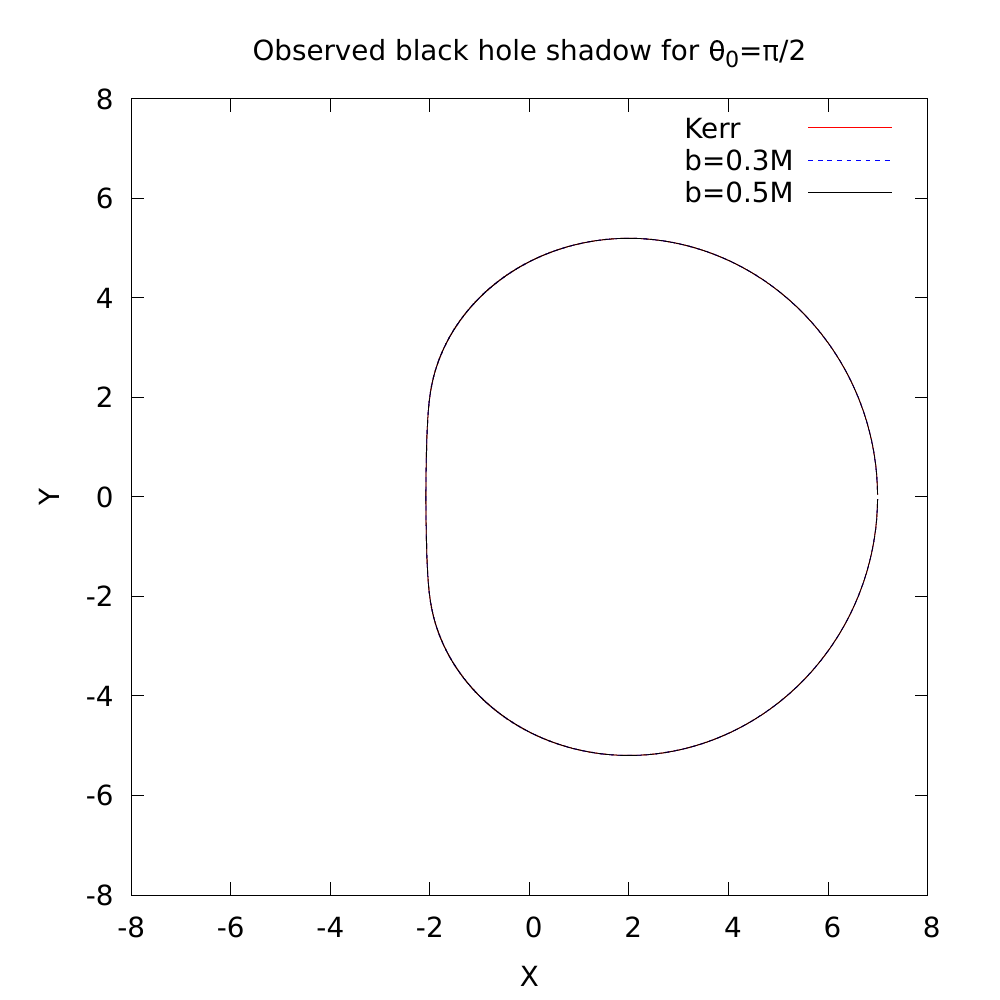}}
\subfigure[Rotating BH without $\mathbb{Z}_2$ symmetry]{\includegraphics[scale=0.8]{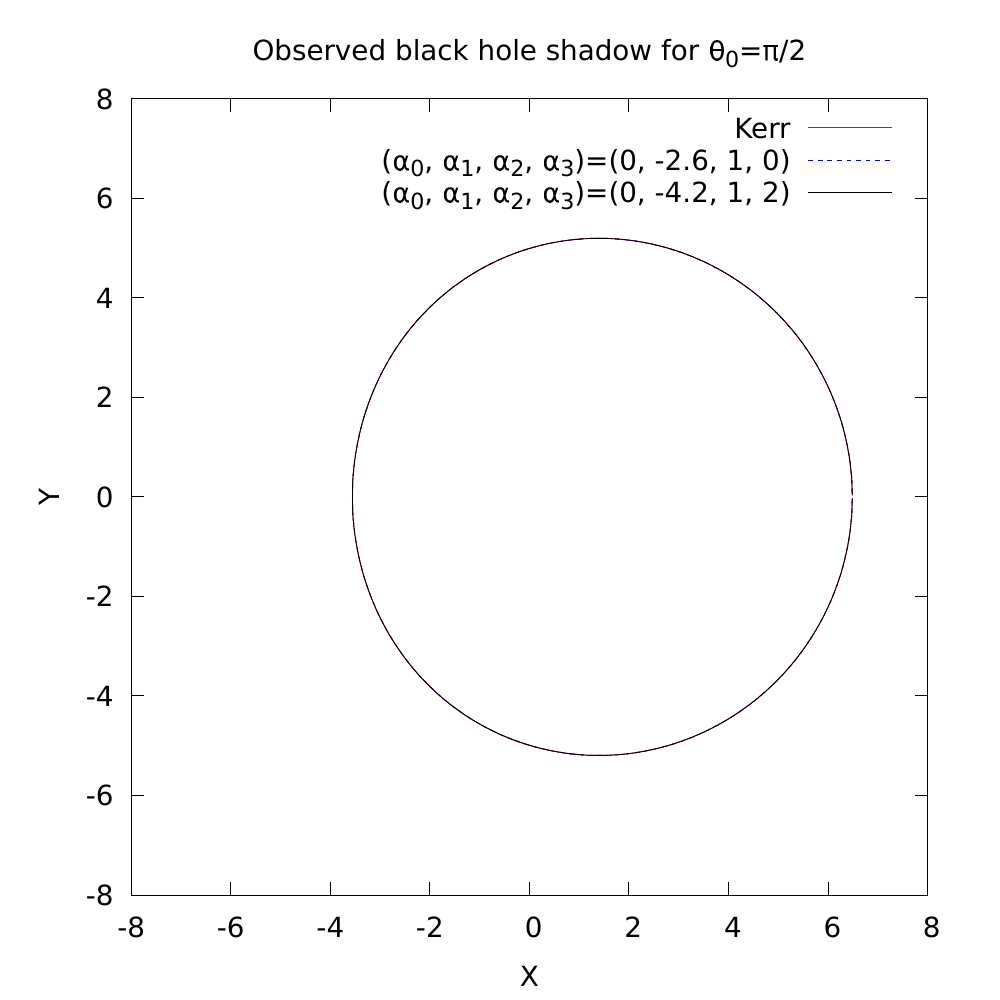}}
\caption{{Panel a: shadow of the rotating SV spacetime, for different values of the parameter $b$ and $a=0.999M$. 
{Panel b: Shadow of a BH without the $\mathbb{Z}_2$ symmetry, for different values of $\alpha_i$ and the same value of the BH spin ($a=0.7M$).
The observer is located at spatial infinity.}}}
\label{figura01}
\end{figure}
\begin{figure}
  \centering
  \subfigure{\includegraphics[scale=0.1]{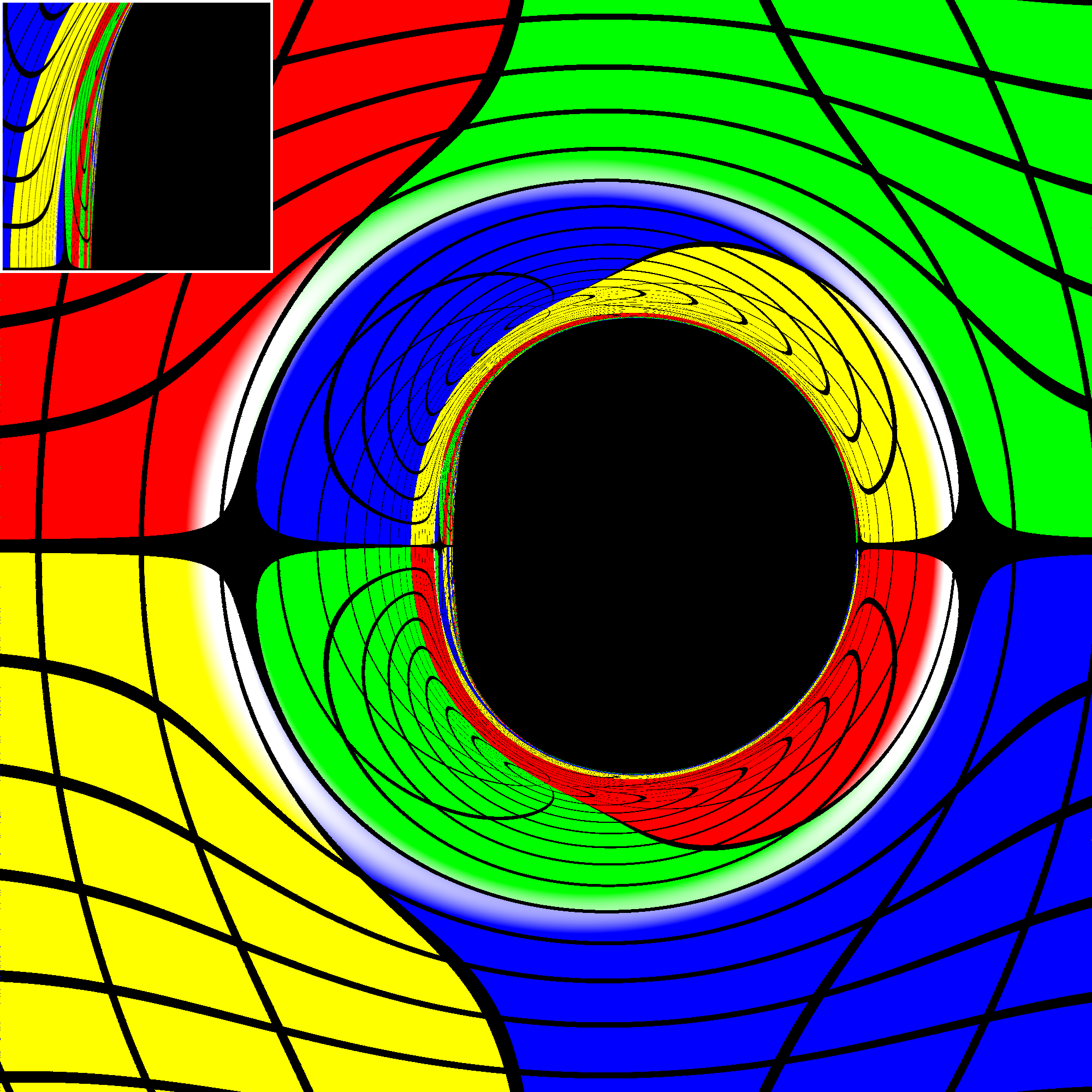}}
  \subfigure{\includegraphics[scale=0.1]{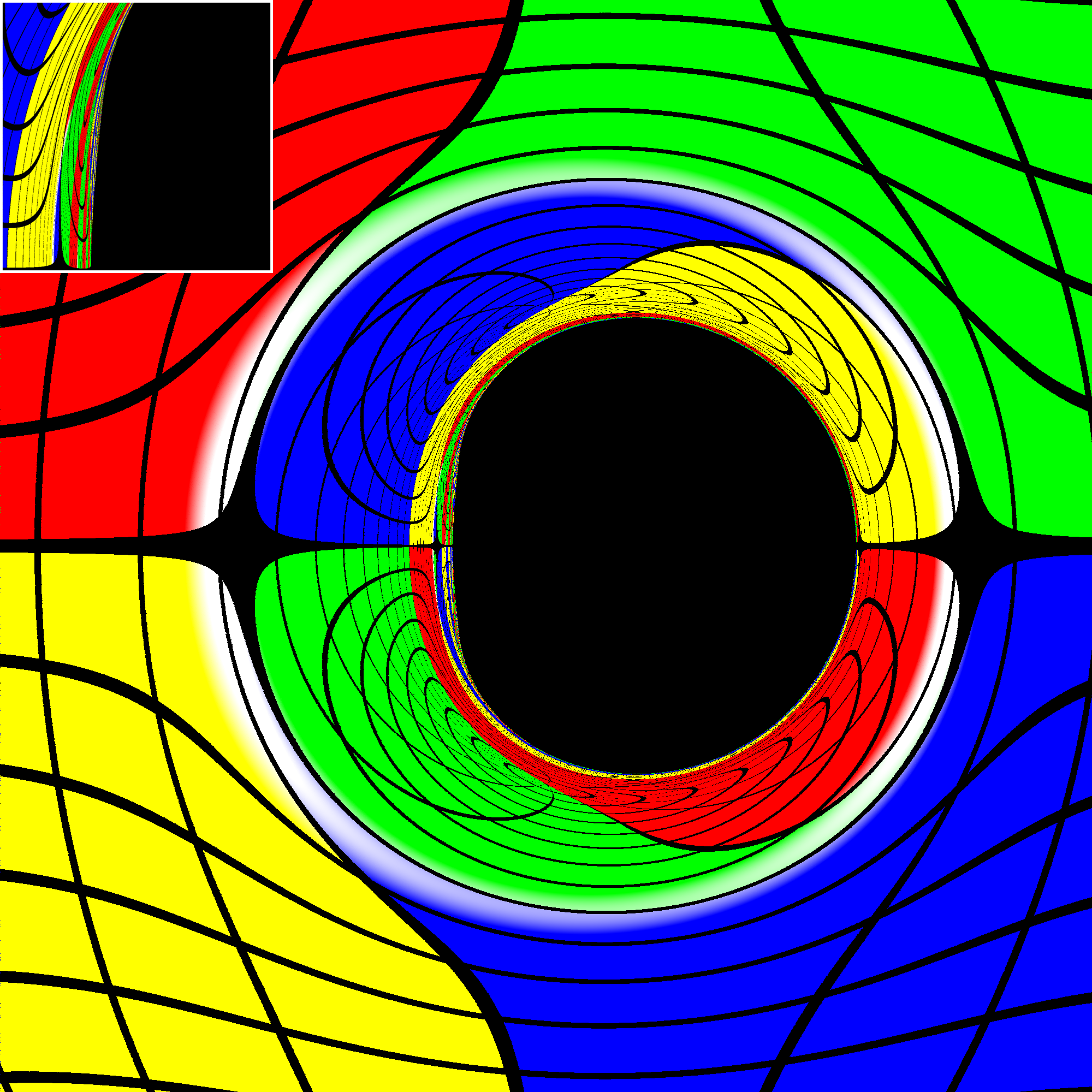}}
  \subfigure{\includegraphics[scale=0.1]{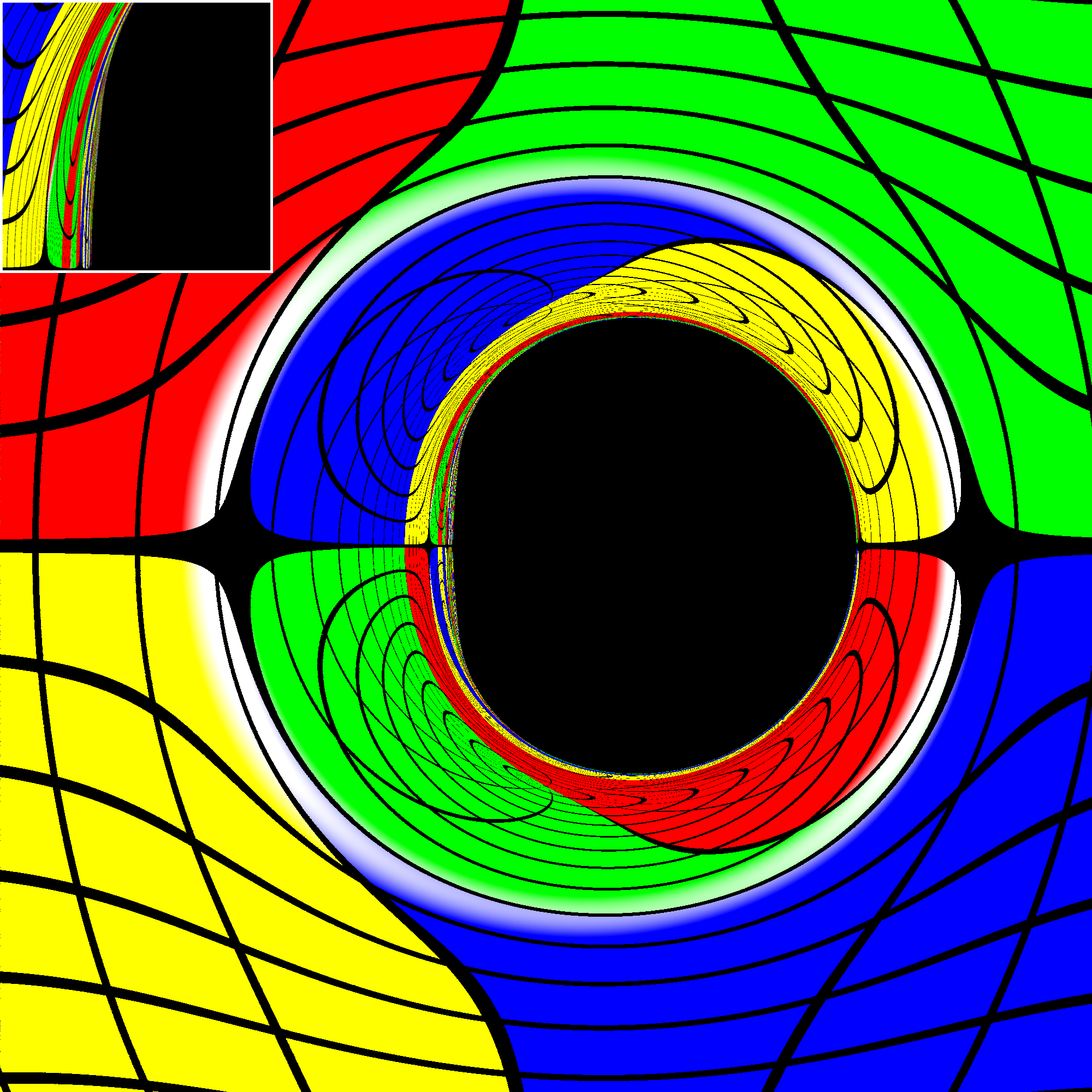}}
  \caption{{The shadow and gravitational lensing for the rotating SV spacetime, for $a=0.999M$ and different values of $b$. In the top panel, we have the Kerr spacetime ($b=0$). In the middle and bottom panels, we have $b=0.5M$ and $b=M$, respectively. The observer position and angle of view are the same as in Figs.~\ref{shadow_SV} and \ref{shadow_CII}. Although the images look the same, there is a subtle difference in the gravitational lensing observed mainly next to the shadow edge.}}
  \label{shadow_NJ_SV}
 \end{figure}
%
%

\subsection{Black holes without $\mathbb{Z}_2$ symmetry}
It has been pointed out that the BH shadow is \textit{not} a probe of the event horizon geometry~\cite{Cunha:2018gql}. 
As a second example of shadow-degeneracy in rotating, stationary and axisymmetric BHs, allowing separability of the HJ equation, we shall provide a sharp example of the previous statement. We shall show that a rotating, stationary and axisymmetric BH without $\mathbb{Z}_2$ symmetry (i.e. without a north-south hemispheres discrete invariance) can, nonetheless, be shadow degenerate with the Kerr BH.

Geometries within the family \eqref{DS_lineel} are  $\mathbb{Z}_2$ symmetric if invariant under the transformation
\begin{equation}
\theta\rightarrow \pi-\theta,
\end{equation}
which maps the north and south hemispheres into one another. The Kerr~\cite{Kerr:1963ud}, Kerr-Newman~\cite{Newman:1965my}, Kerr-Sen~\cite{Sen:1992ua} and the rotating SV spacetimes are examples of BHs with $\mathbb{Z}_2$ symmetry. But BHs without $\mathbb{Z}_2$ symmetry are also known in the literature. One example was constructed in Ref.~\cite{Rasheed:1979}; this property was explicitly discussed in in Ref.~\cite{Cunha:2018prd}, where the corresponding shadows were also studied. 

The general line element displaying shadow degeneracy, under the assumptions discussed above is given in Eq.~\eqref{DS_lineel}. It has a dependence on the $\theta$ coordinate through the function $B_1$ [see Eq.~\eqref{sigma_tilde}].  $B_1(\theta)$ needs not be invariant under $\mathbb{Z}_2$ reflection:
\begin{equation}
B_1(\theta)\neq B_1(\pi-\theta) \ ; 
\end{equation}
then, the BH geometry is shadow degenerate \textit{and} displays no $\mathbb{Z}_2$ symmetry. 

In order to provide a concrete example, $B_1(\theta)$ can be chosen to be
\begin{equation}
\label{B1_NoZ}\cos^2\theta B_1(\theta)=\alpha_0+\alpha_1\cos\theta+\alpha_2\cos^2\theta+\alpha_3\cos^3\theta,
\end{equation}
while $A_1$ and $A_4$ are left unconstrained. In Eq.~\eqref{B1_NoZ}, $\alpha_0$, $\alpha_1$ and $\alpha_2$ are constant parameters for which $\widetilde{\Sigma}>0$. If
\begin{equation}
\alpha_0=\alpha_1=\alpha_3=0, \quad \text{and} \quad \alpha_2=1,
\end{equation}
we recover the Kerr metric (provided that $A_1=A_4=1$). In Fig.~\ref{horizons_noz2}, we show the Euclidean embedding of the event horizon geometry of this BH spacetime for $a/m=0.7$.\footnote{For higher values of the spin parameter $a/m$, it may be impossible to globally embed the spatial sections of the event horizon geometry in Euclidean 3-space. This is also the case for the Kerr spacetime~\cite{Smarr:1973prd} — see also~\cite{Gibbons:2009qe}.} 
We also show the corresponding result for the Kerr BH~\cite{Smarr:1973prd} (top panel). In the middle panel, we have chosen the constants $(\alpha_0,\alpha_1,\alpha_2,\alpha_3)=(0,\ -2.6,\ 1,\ 0)$, while in the bottom panel we have chosen $(\alpha_0,\alpha_1,\alpha_2,\alpha_3)=(0,\ -4.2,\ 1,\ 2)$.  We note that the $\mathbb{Z}_2$ symmetry is clearly violated. Nonetheless, the shadow is always degenerate with the Kerr BH one, as can be seen in the bottom panel of Fig.~\ref{figura01}.

\begin{figure}
  \centering
   \subfigure{\includegraphics[scale=0.27]{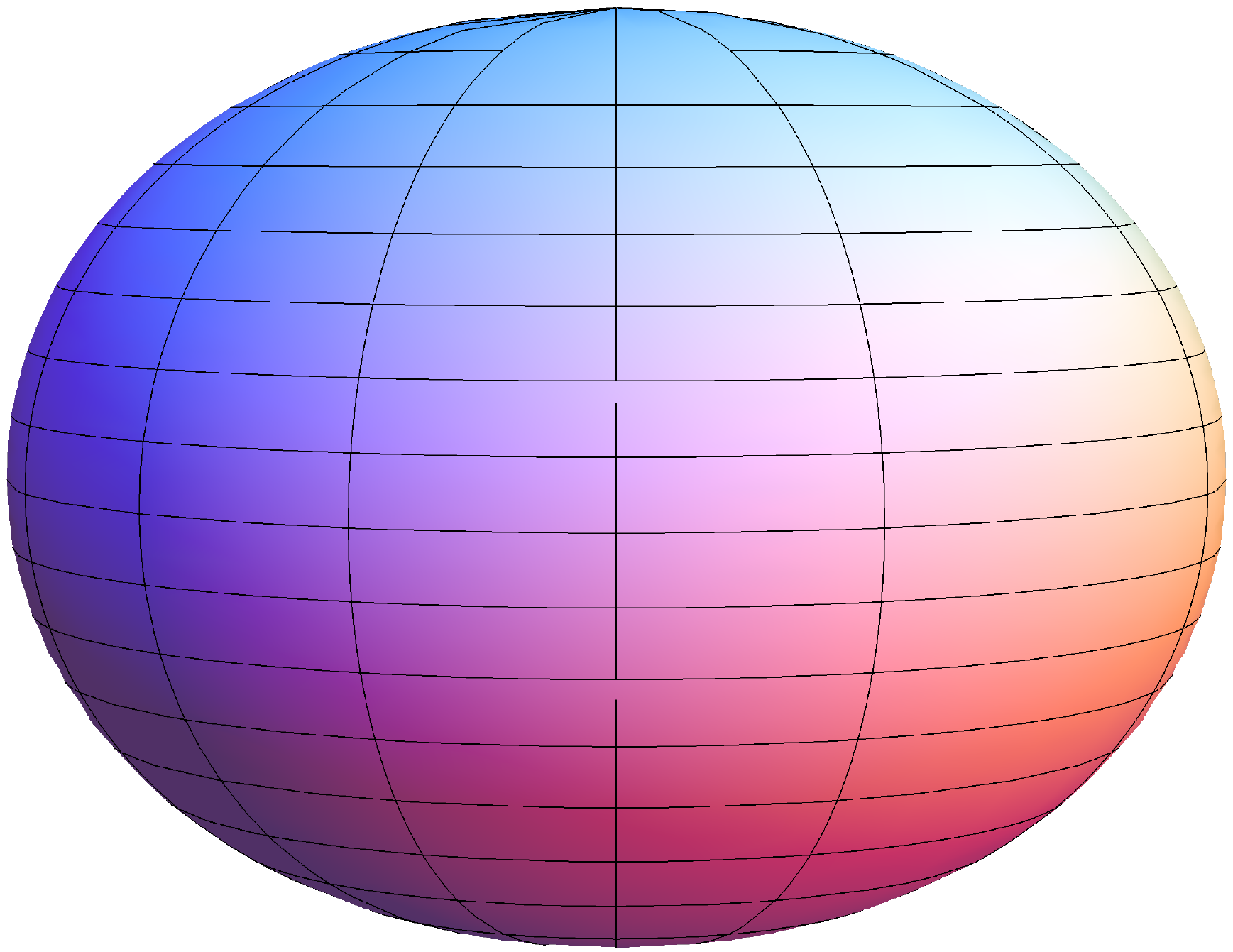}}
  \subfigure{\includegraphics[scale=0.37]{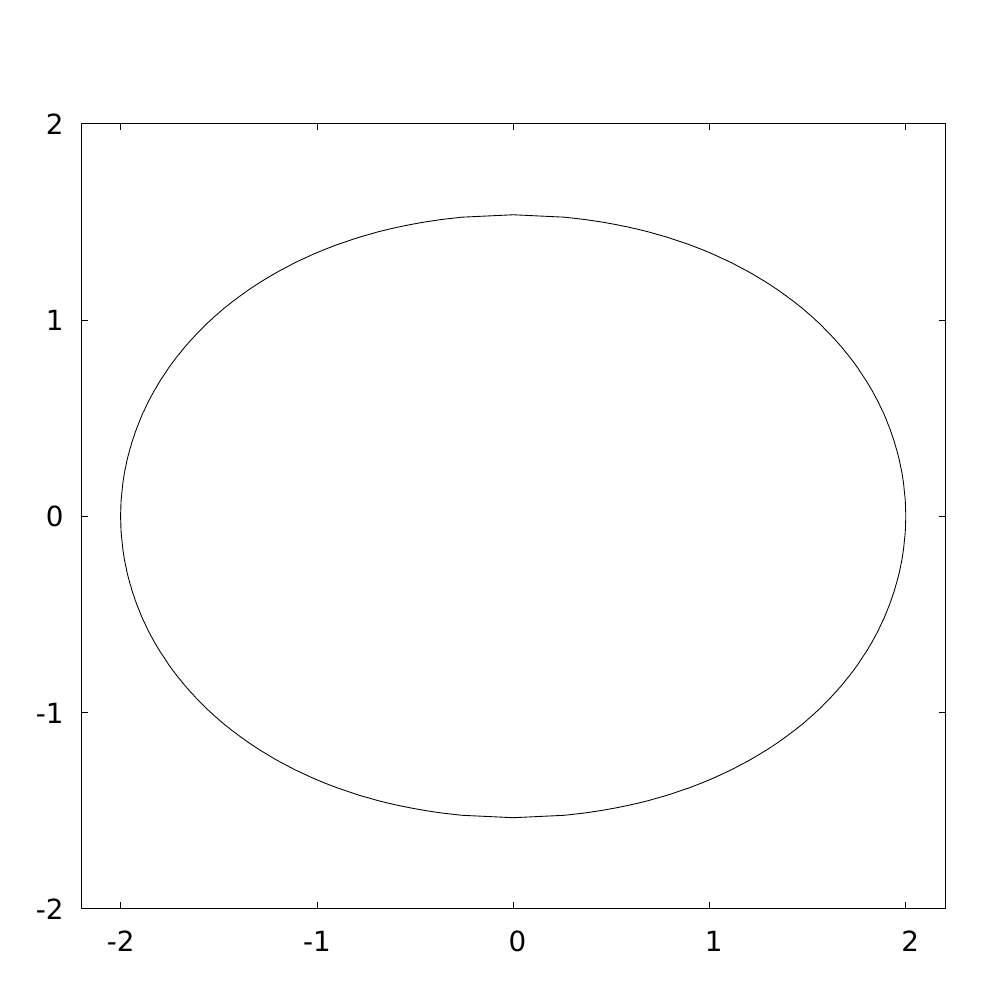}}
   \subfigure{\includegraphics[scale=0.27]{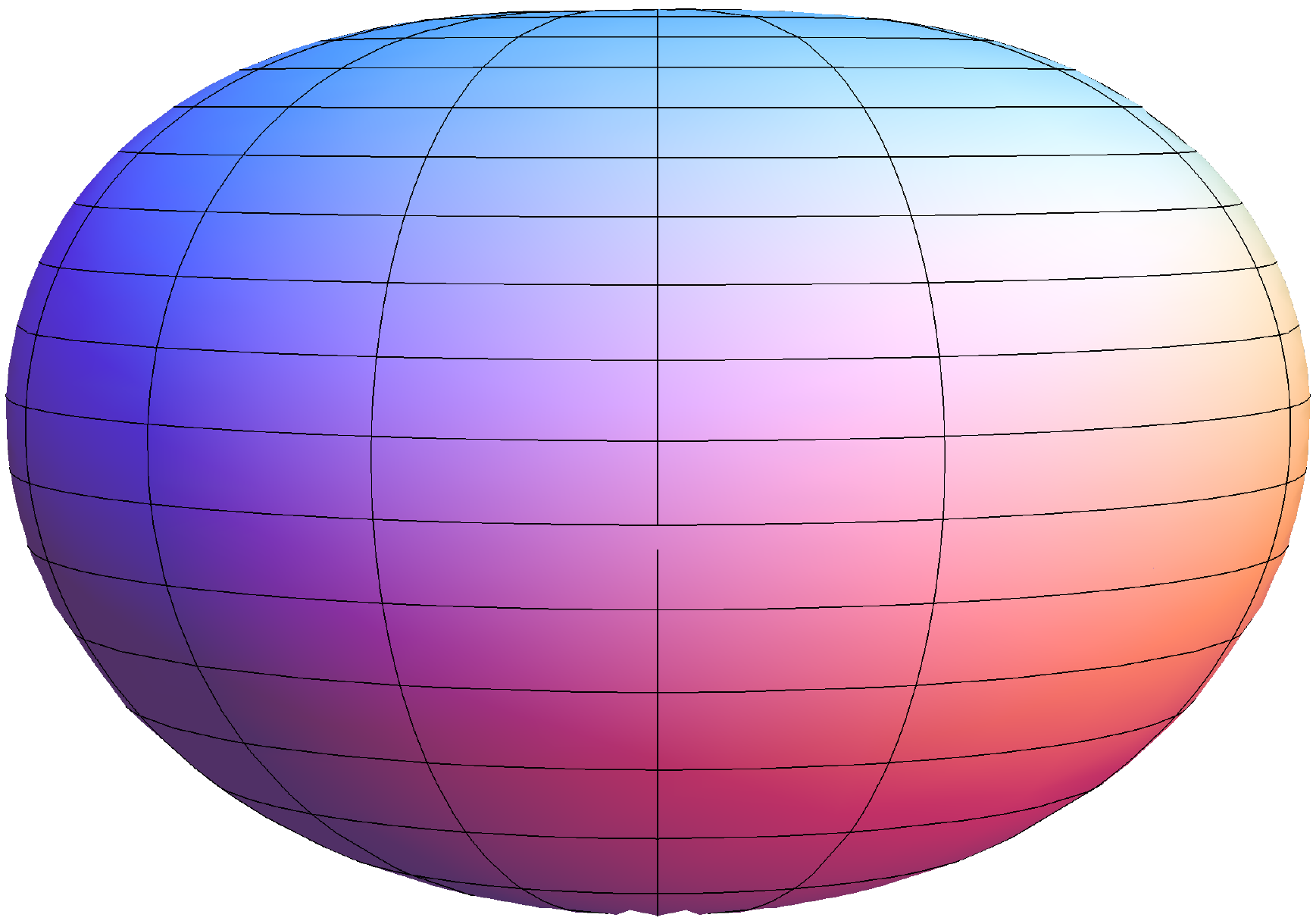}}
  \subfigure{\includegraphics[scale=0.37]{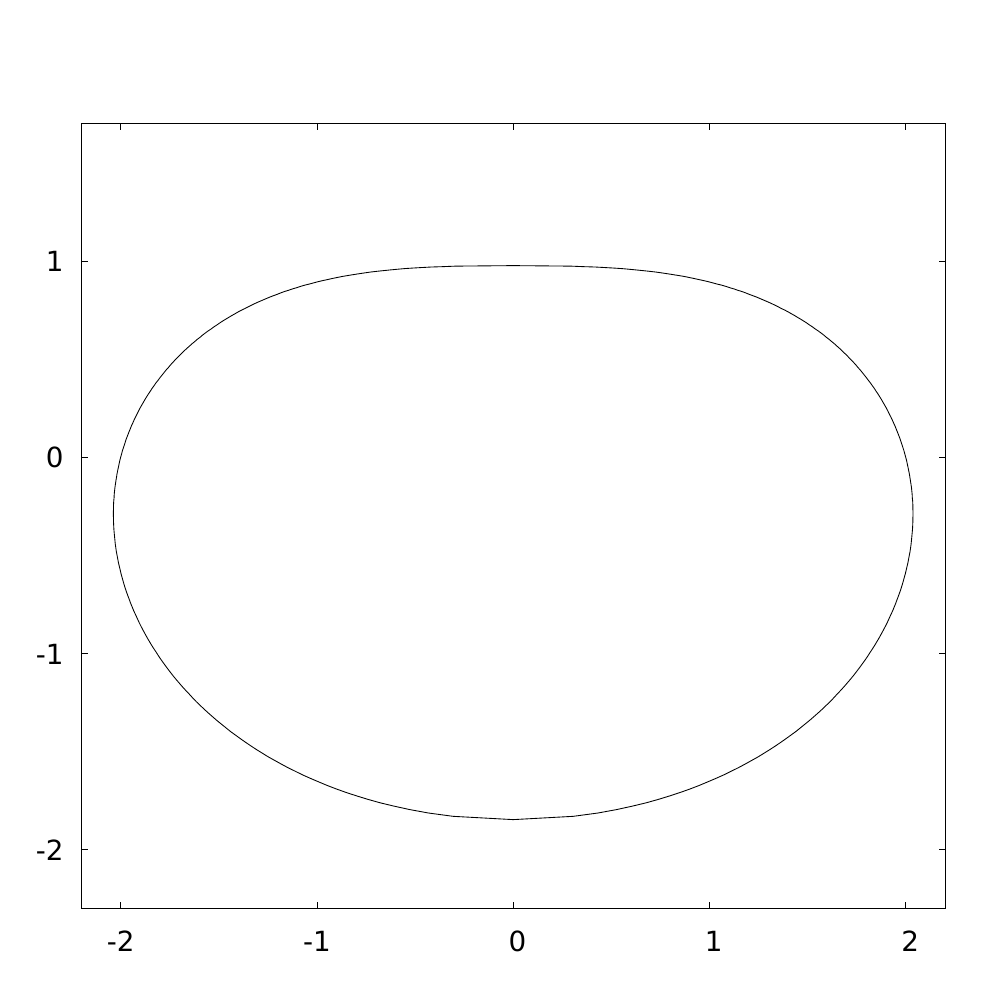}}
   \subfigure{\includegraphics[scale=0.27]{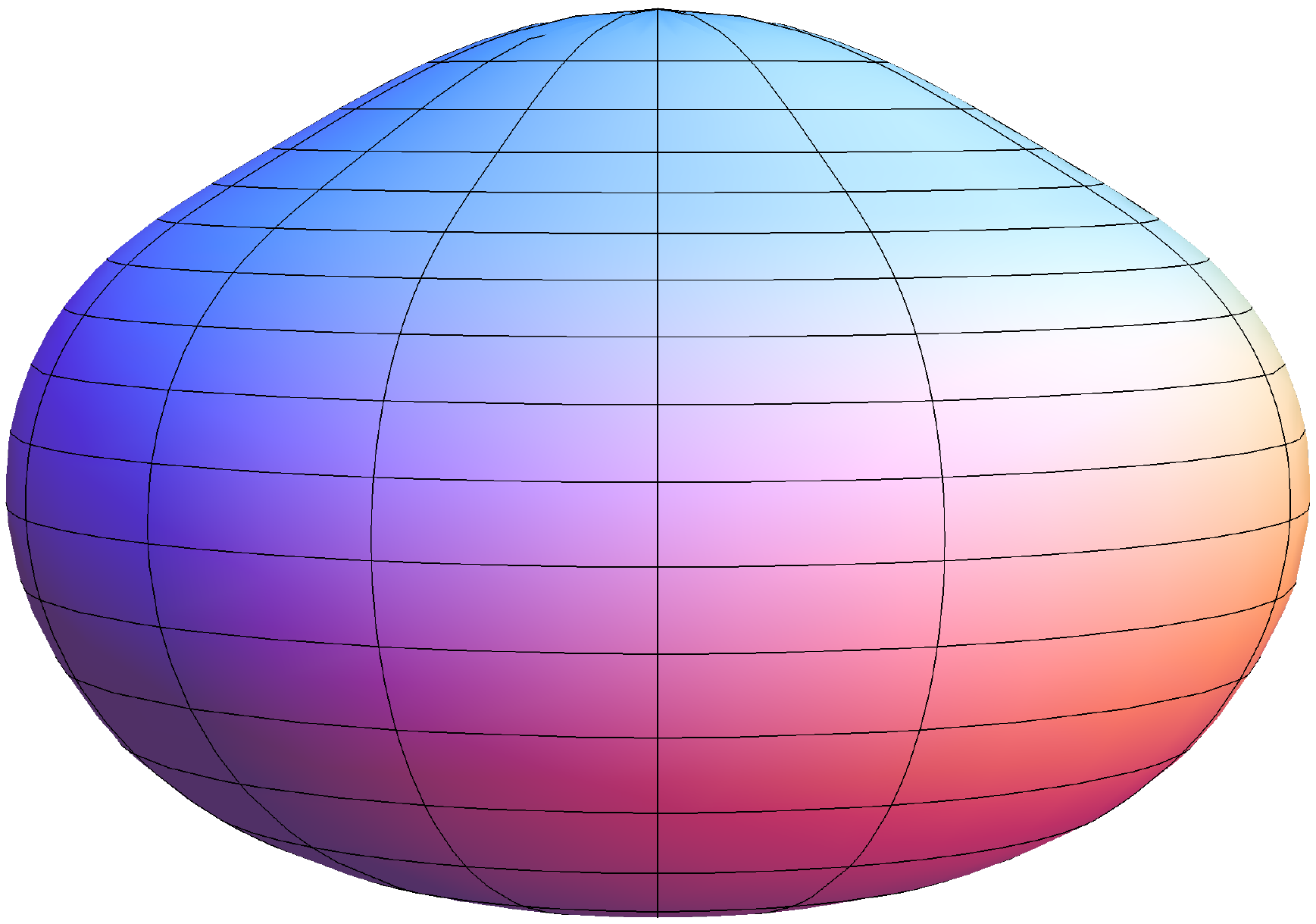}}
  \subfigure{\includegraphics[scale=0.37]{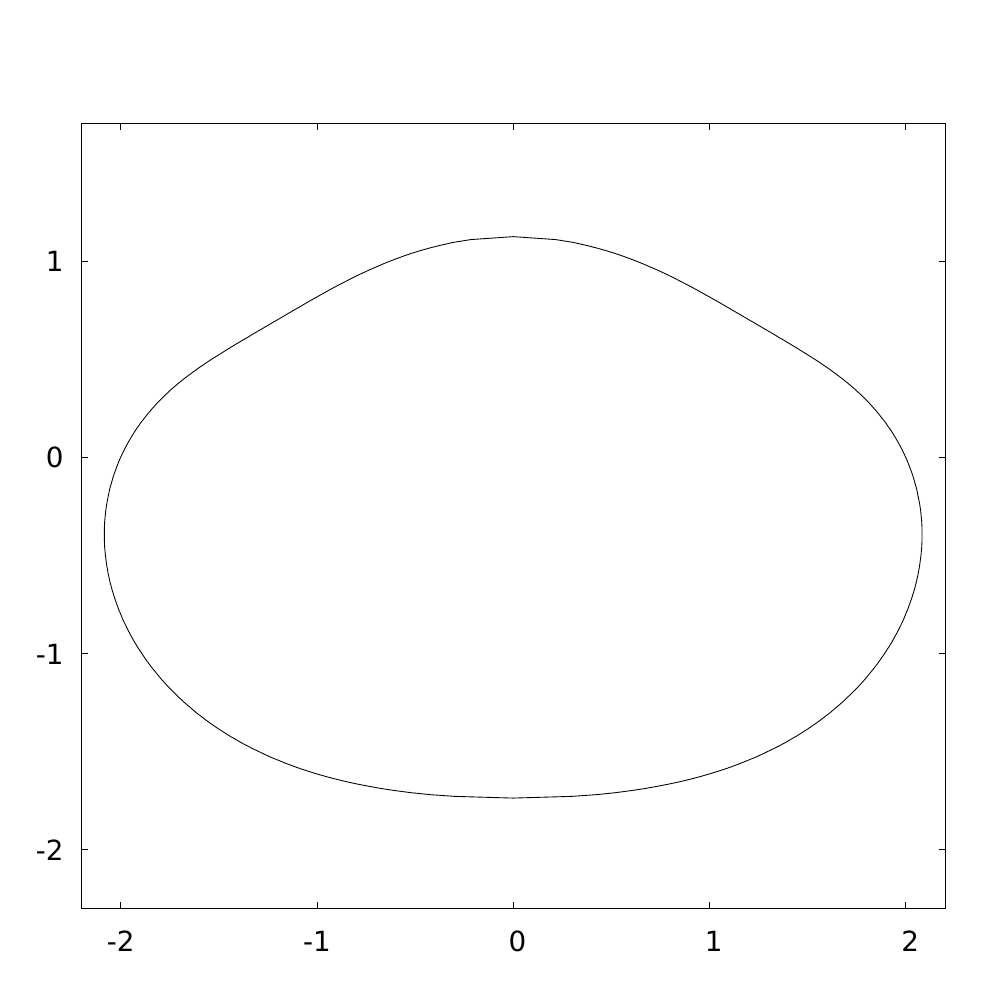}}
  \caption{{Euclidean embedding of the (spatial sections of the) event horizon geometry for the shadow degenerate BHs without $\mathbb{Z}_2$ symmetry. In the top panel, we show the Kerr result, while in the middle and bottom panels we have chosen $(\alpha_0,\alpha_1,\alpha_2,\alpha_3)=(0,\ -2.6,\ 1,\ 0)$ and $(\alpha_0,\alpha_1,\alpha_2,\alpha_3)=(0,\ -4.2,\ 1,\ 2)$, respectively. $a/m=0.7$ for all cases.}}
  \label{horizons_noz2}
 \end{figure}

\section{Conclusions}
\label{sec6}
The imaging of the M87 supermassive BH by the Event Horizon Telescope collaboration~\cite{Akiyama:2019cqa,Akiyama:2019fyp,Akiyama:2019eap} has established the effectiveness of very large baseline interferometry to probe the strong gravity region around astrophysical BHs. In due time, one may expect more BHs, and with smaller masses, will be imaged, increasing the statistical significance of the data. It is then expected these data can be used to test the Kerr hypothesis and even general relativity, by constraining alternative gravity models. It is therefore timely to investigate theoretical issues related to the propagation of light near BHs and, in particular, the properties of BH shadows~\cite{Falcke:1999pj,Cunha:2018acu,Aaron:2019}. 

In this paper, we  investigated the issue of degeneracy: under which circumstances, two different BH spacetimes cast exactly the same shadow? We have established generic  conditions for equilibrium  BHs,  both static and  stationary to be  shadow degenerate with the Schwarzschild or Kerr geometry, albeit in the latter case under the restrictive (but plausible) assumption that precise shadow degeneracy occurs for  spacetimes admitting the separability of the HJ  equation and the same SPO structure as Kerr. We have provided  illustrative examples, in both cases, which highlight, among other properties,  that exact  shadow degeneracy is not,  in general, accompanied  by lensing degeneracy, and that shadow   degeneracy can occur even for qualitatively distinct horizon geometries, for instance, with and without north-south symmetry. 

The examples  herein are mostly of  academic interest,  as a means  to  understand light bending  near compact objects. These examples could be  made  more realistic by considering general relativistic magneto-hydrodynamic simulations  in the BH backgrounds studied here, to   gauge the extent  to which such precise shadow degeneracy survives a more  realistic astrophysical setup. 

Finally, we remark there is also a trivial instance of shadow degeneracy when the same geometry solves different models. This is not the case discussed herein.

\acknowledgements
The authors thank Funda\c{c}\~ao Amaz\^onia de Amparo a Estudos e Pesquisas (FAPESPA),  Conselho Nacional de Desenvolvimento Cient\'ifico e Tecnol\'ogico (CNPq) and Coordena\c{c}\~ao de Aperfei\c{c}oamento de Pessoal de N\'{\i}vel Superior (Capes) - Finance Code 001, in Brazil, for partial financial support.
This work is supported by the Center for Research and Development in Mathematics and Applications (CIDMA) through the Portuguese Foundation for Science and Technology (FCT - Funda\c{c}\~ao para a Ci\^encia e a Tecnologia), references No. BIPD/UI97/7484/2020, No. UIDB/04106/2020 and No. UIDP/04106/2020. We acknowledge support  from the projects No. PTDC/FIS-OUT/28407/2017, No. CERN/FIS-PAR/0027/2019 and No. PTDC/FIS-AST/3041/2020. This work has further been supported by  the  European  Union's  Horizon  2020  research  and  innovation  (RISE) programme No. H2020-MSCA-RISE-2017 Grant No.~FunFiCO-777740. The authors would like to acknowledge networking support by the COST Action No. CA16104.

\bigskip


\bibliography{Ref}

 
\end{document}